\documentclass[aps,preprint,nofootinbib,preprintnumbers,eqsecnum,superscriptaddress]{revtex4-1}

\usepackage{color}

\newcommand{\hl}[1]{\color{magenta}}

\usepackage[
	  pagebackref=false,
	  colorlinks=true,
      linkcolor=blue,
      urlcolor=blue,
      filecolor=black,
      citecolor=red,
      pdfstartview=FitV,
      pdftitle={},
        pdfauthor={},
        pdfsubject={},
        pdfkeywords={},
        pdfpagemode=None,
        bookmarksopen=true
      ]{hyperref}

\usepackage[normalem]{ulem}
\usepackage{amsmath}
\usepackage{enumerate}
\usepackage{amsfonts}
\usepackage{epsfig}
\usepackage{mathbbol}

\def\Tr{\mathop{\rm Tr}}

\newcommand\half{{\ensuremath{\frac{1}{2}}}}
\newcommand\p{\ensuremath{\partial}}

\newcommand\field[1]{{\ensuremath{\mathbb{{#1}}}}}

\newcommand\vev[1]{{\ensuremath{\left\langle{#1}\right\rangle}}}

\newcommand{\ZZ}{\field{Z}}

\newcommand{\be}{\begin{equation}}
\newcommand{\ee}{\end{equation}}
\newcommand{\bea}{\begin{eqnarray}}
\newcommand{\eea}{\end{eqnarray}}
\newcommand{\bega}{\begin{gather}}
\newcommand{\eega}{\end{gather}}

\newcommand{\bi}{\begin{itemize}}
\newcommand{\ei}{\end{itemize}}
\newcommand{\ben}{\begin{enumerate}}
\newcommand{\een}{\end{enumerate}}
\newcommand{\bca}{\begin{cases}}
\newcommand{\eca}{\end{cases}}
\newcommand{\bln}{\begin{align}}
\newcommand{\eln}{\end{align}}
\newcommand{\bst}{\begin{split}}
\newcommand{\est}{\end{split}}
\def\ie{\begin{equation}\begin{aligned}}
\def\fe{\end{aligned}\end{equation}}
\newcommand{\bma}{\le(\begin{matrix}}
\newcommand{\ema}{\end{matrix}\ri)}

\newcommand\al{{\alpha}}
\newcommand\ep{\epsilon}
\newcommand\sig{\sigma}
\newcommand\Sig{\Sigma}
\newcommand\lam{\lambda}
\newcommand\Lam{\Lambda}
\newcommand\om{\omega}
\newcommand\Om{\Omega}

\newcommand\ga{{\ensuremath{{\gamma}}}}

\newcommand\de{{\ensuremath{{\delta}}}}
\newcommand\De{{\ensuremath{{\Delta}}}}
\newcommand\vp{\varphi}

\newcommand\vep{\varepsilon}
\newcommand\Th{{\Theta}}
\def\th{{\theta}}

\newcommand\da{{\dagger}}
\newcommand\nab{{\nabla}}

\newcommand\wg{\wedge}

\newcommand\ov{\over}
\newcommand\ha{{\half}}

\def\le{\left}
\def\ri{\right}

\newcommand\sB{{\ensuremath{{\mathcal B}}}}
\newcommand\sC{{\ensuremath{{\mathcal C}}}}

\newcommand\sF{{\ensuremath{{\mathcal F}}}}

\newcommand\sL{{\ensuremath{{\mathcal L}}}}

\newcommand\sO{{\ensuremath{{\mathcal O}}}}
\newcommand\sP{{\ensuremath{{\mathcal P}}}}

\newcommand\sJ{{\mathcal J}}

\newcommand\sT{{\mathcal T}}

\newcommand\vx{{\vec x}}

\newcommand{\hmu}{{\hat \mu}}

\newcommand{\fw}{{\mathfrak w}}
\newcommand{\fB}{{\mathfrak B}}

\begin{document}

\title{Global Anomalies,  Discrete Symmetries, and Hydrodynamic
Effective Actions}

\preprint{MIT-CTP/4944}

\author{Paolo Glorioso}
\affiliation{Kadanoff Center for Theoretical Physics and Enrico Fermi Institute\\
University of Chicago, Chicago, IL 60637, USA}

\author{Hong Liu}
\affiliation{Center for Theoretical Physics, \\
Massachusetts
Institute of Technology,
Cambridge, MA 02139 }

\author{Srivatsan Rajagopal}
\affiliation{Center for Theoretical Physics, \\
Massachusetts
Institute of Technology,
Cambridge, MA 02139 }

\begin{abstract}

\noindent

We derive effective actions for  parity-violating fluids in both $(3+1)$ and $(2+1)$ dimensions, including those with
anomalies. As {a} corollary we confirm the  most general constitutive relations for such systems derived {previously} using other methods. We discuss in detail connections between parity-odd transport and underlying discrete symmetries.
In (3+1) dimensions we elucidate connections between anomalous transport coefficients and global anomalies,
and clarify a previous puzzle concerning transports and local gravitational anomalies.


\end{abstract}

\today

\maketitle

\tableofcontents

\section{Introduction}

Through studies of free field theories~\cite{Vilenkin:1979ui,Vilenkin:1980zv,Vilenkin:1980fu,Volovik:1999wx}, holographic duality~\cite{Erdmenger:2008rm,Banerjee:2008th,Loganim:2013}, phenomenological arguments based on
entropy current~\cite{Son:2009tf,Neiman:2010,Loganim:2011,Andrey:2010,Andrey:2010a}, and equilibrium partition functions~\cite{Banerjee:2012iz,Jensen:2012jh,Loganim:2012}, it has been recognized that
systems with quantum anomalies exhibit novel transport behavior in
the presence of rotation or in a magnetic field (for a recent review see~\cite{Landsteiner:2016led}).
Effects of anomalies on transport in superfluids, superconductors and topological insulators have also been discussed in~\cite{Neiman:2011mj,Bhattacharya:2011tra,Ryu:2012aa,Stone:2012}. (See also~\cite{Kharzeev:2004ey,Kharzeev:2007tn,Sone:2004,metlitski:2005,Torabian:2009qk,Jensen:anom,Kharzeev:2014,Valle:2012}.)
These anomalous transports could be  relevant in a wide range of physical contexts: from the study of quark-gluon plasma at subnuclear scales \cite{Kharzeev:2007jp,Fukushima:2008xe,Kharexp:2017,Gang:2017,Kharzeev:2010gr}, to cosmology, where the dynamics of primordial magnetic fields plays an important role in the early stage of the universe \cite{Joyce:1997uy,Giova1}, {and astrophysical phenomena such as pulsar kicks~\cite{Kaminski:2014jda,Shaverin:2014xya}.} In addition, there have been various experimental searches for the signatures of anomalies on transports 
in condensed matter systems, see \cite{Qiang:2014, Gooth:2017,Shekhar:2015rqa}.

Given their importance, it is of primary interest to incorporate anomalous transports in an
effective field theory framework, which is the goal of this paper. Such a formulation has a number of
advantages. Firstly, an effective field theory provides a framework where hydrodynamic fluctuations can be systematically incorporated, thus enabling one to search for new physical effects due to fluctuations in parity-violating systems.
Secondly, 
the effective action approach provides a first-principle derivation of the constitutive relations which automatically incorporates
all the phenomenological constraints.
Indeed our derivation reproduces fully the constitutive relations of previous
approaches. It also highlights  some new insights which we will discuss momentarily. 

Consider a parity-violating relativistic system in $(3+1)$-dimension with a global $U(1)$ symmetry whose conserved current is $\hat J^\mu$. Suppose the symmetry becomes anomalous  in the presence of an external source $A_\mu$ for $\hat J^\mu$,
 \be
 \nab_\mu \hat J^\mu = {c \hbar \ov 4}  \ep^{\mu \nu \al \beta} F_{\mu \nu} F_{\al \beta} 
 \label{AEq1}
 \ee
where $F$ is the field strength for $A$.  Due to~\eqref{AEq1}, the Euclidean partition function of the system
in the presence of source $A_\mu$ is not invariant under {\it small} gauge transformations
of $A$.
We will refer to~\eqref{AEq1} as
a local $U(1)$ anomaly,  in contrast to a global anomaly in which case the partition function is invariant
under small gauge transformations, but not under {\it large} gauge transformations when the system is put
on a topologically nontrivial manifold.
To first order in the derivative expansion, the parity-odd part $J_o^\mu$ of  charge
current  
can be written in the Landau frame as~\cite{Son:2009tf,Neiman:2010,Kharzeev:2007jp,Fukushima:2008xe}
\be
J^\mu_o = 
 \xi_\om \om^\mu + \xi_B\, \sB^\mu \ .
\label{vortcurrent}
 \ee
The first term implies a contribution to the current that is induced by and parallel to,
the vorticity {$\om^\mu \equiv  \ep^{\mu \nu \lam \rho} u_\nu \p_\lam u_\rho$} ($u^\mu$ is the local velocity
field). This is called the chiral vortical effect (CVE).
The second term is proportional to the
magnetic field strength $\sB^\mu\equiv \frac{1}{2}\epsilon^{\mu\nu\alpha\beta}u_\nu F_{\alpha\beta}$,
which is often referred to as the chiral magnetic effect (CME). 
The transport coefficients $\xi_\om$ and $\xi_B$ receive contributions from local anomaly~\eqref{AEq1}
as follows~\cite{Son:2009tf,Neiman:2010,Banerjee:2012iz}
\bea \label{CVE1}
\xi_\om & =& -3 c \hbar \mu^2 \le(1 - {2 \ov 3} \al \ri) + 2 a_1 \mu T \le(1 - \al \ri)
+ a_2 T^2 \le(1 - 2  \al \ri) - 2 a_3 T^3 {n_0 \ov  \ep_0 + p_0}  , \\
\xi_B & =& -3 c \hbar \mu   (2 - \al) + 2 a_1 T (1 - \al)  - a_2 T^2   {n_0 \ov  \ep_0 + p_0}  ,
\label{CME1}
\eea
with
\be \al \equiv  {\mu n_0 \ov  \ep_0 + p_0}
\ee
 where $a_{1,2,3}$ are constants, and $\mu, T, n_0, \ep_0, p_0$ are local chemical
 potential,  temperature, charge density, energy density and pressure respectively.

 It is curious that even in the absence of local anomaly~\eqref{AEq1}, i.e. with $c=0$,
 there can still be chiral vortical and magnetic effects, determined up to three
 constants. 
 It has been pointed out that for a CTP invariant theory, only $a_2$ is allowed~\cite{Bhattacharya:2011tra,Banerjee:2012iz},
 whose physical origin has generated much recent interest. 
 From holography and free theory examples, $a_2$ appears to be related to the coefficient $\lam$ of
the local mixed gravitational anomalies
 \be
 \nab_\mu \hat J^\mu = \lam \epsilon^{\mu\nu\lam\rho} R^{\alpha}{_{\beta\mu\nu}} R^{\beta}{_{\alpha\lam\rho}} \
 \label{AEq2}
 \ee
as~\cite{Landsteiner:2011iq,Landsteiner:2011cp,Landsteiner:2012kd,Loganayagam:2012pz,Chowdhury:2015pba,Andrey:2017,Chapman:2012}
\be \label{mat}
a_2=-32 \pi^2\lambda  \ .
\ee
{Relation~\eqref{mat} is puzzling from the perspective of anomaly matching in a low energy effective theory}, as the right hand side of~\eqref{AEq2} contains four derivatives and thus should modify $J^\mu$ only at the third derivative order while terms in~\eqref{vortcurrent} have only one derivative.
Furthermore, matching with constitutive relations or partition functions as done in~\cite{Son:2009tf,Neiman:2010,Banerjee:2012iz,Jensen:2012jh} will not lead to any multiplicative factor $\pi$ as in~\eqref{mat}.
{Arguments have been made in~\cite{Jensen:2012kj,Jensen:2013rga,Jensen:2013kka,DiPietro:2014bca,Majhi:2014} which show that~\eqref{mat} should apply at least to field theory systems smoothly connected to free theories through continuous parameter(s).}  Alternatively, it has been hinted in~\cite{Golkar:2012kb} and subsequently explicitly worked out in various examples
in~\cite{Golkar:2015oxw,Chowdhury:2016cmh} that the transport coefficient $a_2$
should be considered as being directly related to global mixed gravitational anomalies when putting the system on a topologically nontrivial manifold. It has also been known that relation like~\eqref{mat} is violated for systems with gravitinos~\cite{Chowdhury:2015pba,Logan22:2012,Jensen:2012kj,Jensen:2013kka}.


In this paper we work out  effective actions for parity-violating fluids in both $(2+1)$ and $(3+1)$ dimensions  following the approach developed in~\cite{CGL,CGL1,GL} (see~\cite{Dubovsky:2011sk,Haehl:2013hoa,Monteiro:2014wsa} for earlier attempts at an effective action for anomalous transports). 
We assume that at microscopic level the system has
an underlying discrete symmetry $\Th$ which includes time reversal. Here  $\Th$ can be the time reversal $\sT$ itself, or any combinations of $\sC, \sP$ with $\sT$, such as $\sC\sP\sT$. As a corollary we confirm~\eqref{vortcurrent}--\eqref{CME1}
as the most general constitutive relation for a parity-violating system in $(3+1)$-dimensions, and in $(2+1)$-dimension we confirm the constitutive relations obtained earlier in~\cite{Jensen:2011xb,Banerjee:2012iz,Jensen:2012jh}.
In $(2+1)$-dimension the story is much richer, containing six independent functions of local temperature and chemical potential.
The rest of the paper is devoted to detailed derivations of the effective actions. Here we highlight a
couple of conceptual points related to~\eqref{CVE1}--\eqref{CME1}. {In particular, we offer an interpretation for~\eqref{mat}
which reconciles various different perspectives.}\footnote{While these points follow naturally from our discussion, some aspects could have been realized before using the approaches already discussed in the literature. For example, the connection with global anomalies discussed below could have been read from the results of~\cite{Banerjee:2012iz}.} {We find:} 

\ben

\item In both $(3+1)$ and $(2+1)$ dimensions, possible parity-odd transport behavior sensitively depends on the underlying discrete
symmetries. Hence hydrodynamic transports can be used to probe microscopic discrete symmetries.
For example, given the form~\eqref{vortcurrent}--\eqref{CME1}, when $\sP \sT$ is conserved, then $a_{1,2,3} =0$ and $c=0$, i.e. no chiral vortical or magnetic
effects. If $\sC \sP \sT$ is conserved, then $a_1 = a_3 =0$. 
If only $\sT$ is conserved, then all $a_{1,2,3}$ and $c$ are allowed. {Thus detection of possible existence of $a_1, a_3$ can be used to test $\sC \sP \sT$ violations.}

While $\sC \sT \sP$ is preserved for all relativistic local field theories, searching for its possible violations
through transports could be interesting. Some condensed matter systems exhibit emergent relativistic symmetries, and
transport behavior can then be potentially used to probe whether there is emergent $\sC \sP \sT$ as well.

\item All three constants $a_{1,2,3}$ in~\eqref{CVE1}--\eqref{CME1} are associated with global anomalies, respectively with pure gauge, mixed gauge, and pure gravitational anomalies.
More explicitly, consider the partition function of the system on a spatial manifold $S^1 \times S^2$ at a finite temperature, i.e. the full manifold is $S_T^1 \times S^1 \times S^2$, with $S_T^1$ denoting the Euclidean time direction along which we put thermal boundary conditions. We also turn on the external metric and  source $A_\mu$ as
\be \label{Bni}
ds^2 = g_{00} \le(d\tau - v_i dx^i \ri)^2 + a_{ij} dx^i dx^j , \qquad A_\mu dx^\mu  = A_0 (d\tau - v_i dx^i)  + b_i dx^i  \
\ee
with all components to be independent of Euclidean time $\tau$. $x^i$ denotes directions along $S^2 \times S^1$.
Let us suppose there is no local gauge anomaly~\eqref{AEq1}, i.e. $c=0$. Then to first derivative order, the partition function
should be invariant under the following two $U(1)$ transformations\footnote{The gravitational anomaly~\eqref{AEq2} does not matter at this derivative order.}
\bega \label{tu12}
v_i \to v_i - \p_i f , \qquad b_i \to b_i , \\
v_i \to v_i, \qquad b_i \to b_i + \p_i g  \
\label{tu22}
\end{gather}
where both $f$ and $g$ are independent of $\tau$. Equation~\eqref{tu12} aries from time diffeomorphism along the Euclidean time circle\footnote{{The transformation associated to $f$ is also known in the literature as Kaluza-Klein $U(1)$.}} while~\eqref{tu22} is the stationary gauge transformation for $A_\mu$. It turns out, however, when $a_{1,2,3}$ are nonzero, the partition function is only invariant under transformations which are smoothly connected to the identity, but not invariant
under large gauge transformations.

More explicitly, suppose $b_i$ has a magnetic flux along $S^2$, then under a large gauge transformation of $b_i$ and $v_i$ along $S^1$ we find that the partition function transforms as
\be \label{ghh1}
{Z \to \exp \le[  {8 \pi^2 m  a_1 \ov q^2} + i {2 \pi   n a_2  \ov q} \ri] Z  , \qquad m,n \in \ZZ}
\ee
where $q$ is the minimal $U(1)$ charge of the system. {The term proportional to $a_2$ in~\eqref{ghh1} is fully consistent with the discussion of various examples in~\cite{Golkar:2015oxw,Chowdhury:2016cmh}.} In~\eqref{ghh1}  the term in the exponent proportional to $a_1$ is real; recall that the
presence of $a_1$ breaks $\sC \sP \sT$. Similarly when only {$v_i$} has a magnetic flux along $S^2$,  under a large gauge transformation of {$v_i$} along $S^1$ we find that
\be \label{ghh2}
Z \to e^{-2a_3 r } Z, \qquad r \in \ZZ \
\ee
which is again real. The standard lore is that there can be no
pure global gravitational anomaly in $d=4$. But here $\sC \sP \sT$ is broken and we are at a finite temperature.

We thus see measuring parity-violating transports can also be used to probe global anomalies of a system.
{Note that $a_2$ appears in~\eqref{ghh1} in a phase, so the global anomaly~\eqref{ghh1} only captures the ``fractional'' part of $a_2$, i.e. $a_2 \to a_2 + k q$ with $k \in \ZZ$ does not change the phase. In contrast, the factors associated with $a_1$ and $a_3$ in~\eqref{ghh1}--\eqref{ghh2} are real. As a result the global anomalies associated with them are fully equivalent to the corresponding transport coefficients.}

{The relations between coefficients $a_{1,2,3}$ and global anomalies described above are universal relations which 
can be deduced solely at the level of low energy effective theory, without any knowledge of UV physics. 
Now let us come back to the relation~\eqref{mat} which from the light of the above discussion may be interpreted as 
the combination of the following:}
\ben
\item  the connection between $a_2$-related transports in~\eqref{CVE1}--\eqref{CME1} to global gravitational anomaly~\eqref{ghh1} {which is a universal low energy relation;}

\item a relation between local mixed anomaly coefficient $\lam$ in~\eqref{AEq2} and the global mixed anomaly~\eqref{ghh1}
which has been known to be valid for some class of systems. {This relation goes beyond low energy physics.} 
\een
{This resolves the two puzzles mentioned below~\eqref{mat}: equation~\eqref{mat} should not be viewed as a low energy relation. Indeed, from the perspective of low energy effective field theory, neither transport behavior in~\eqref{CVE1}--\eqref{CME1} nor the global anomaly in~\eqref{ghh1} has anything to do with~\eqref{AEq2}. Nevertheless, when UV physics is taken into consideration, they are controlled by the same number in a large class of systems. In this light the discussion of~\cite{Jensen:2012kj,Jensen:2013rga,Jensen:2013kka,DiPietro:2014bca} can be considered as establishing (b) for field theory systems smoothly connected to free theories through continuous parameter(s).}

\een

The plan of the paper is as follows. In Sec.~\ref{sec:rev} we briefly review the formalism of~\cite{CGL,CGL1,GL} to set up the notations and the rules for derivations of later sections. In Sec.~\ref{sec:31} we obtain the effective action of a parity-violating fluid in $(3+1)$-dimension. In Sec.~\ref{sec:par} we discuss the connection between the effective action and thermal partition function, and connection with global anomalies. In Sec.~\ref{sec:ent} we discuss the entropy current for $(3+1)$-systems.
In Sec.~\ref{sec:21} we repeat the analysis for $(2+1)$-dimensional parity-violating systems, obtaining the effective action, partition function and the entropy current. We have also included a number of Appendices for technical details.


 \section{Review of hydrodynamical action in physical spacetime} \label{sec:rev}

In this section, we review the formulation of {the} hydrodynamical action introduced in~\cite{CGL,CGL1,GL} to set up the notations and formalism for deriving anomalous transports in later sections.\footnote{See also~\cite{Dubovsky:2005xd,Dubovsky:2011sj,Endlich:2012vt,Grozdanov:2013dba,Kovtun:2014hpa,Harder:2015nxa,Haehl:2015pja,Haehl:2015uoc,Andersson:2013jga,Floerchinger:2016gtl,yarom,ping} for other discussions of action formulation.}

\subsection{General setup}

Consider the closed time path (CTP)
generating functional $W[ g_1, A_1;  g_2, A_2]$ for a system with a $U(1)$ symmetry in some state specified by {the} density matrix $\rho_0$ 
\be \label{pager1}
e^{W  [ g_{1\mu \nu} , A_{1\mu}; g_{2 \mu \nu}, A_{2 \mu}] }
\equiv   \Tr \le[U (+\infty, -\infty; g_{1\mu \nu}, A_{1\mu}) \rho_0 U^\da (+\infty, -\infty; g_{2\mu \nu}, A_{2 \mu})\ri] 
\ee
where $U(t_2, t_1; g_{1\mu \nu}, A_{1\mu})$ denotes the quantum evolution operator
of the system from $t_1$ to $t_2$
in the presence of spacetime metric $g_{1\mu \nu}$ and an external vector field $A_{1\mu}$ (sources for the $U(1)$ current).
The sources for two legs of the CTP contour are taken to be independent.
We introduce the ``on-shell'' stress tensors and currents for each leg as
\bega \label{defst1}
- i {\de W \ov \de g_{1\mu \nu} (x)}=
 \ha \sqrt{-g_1}  T^{\mu \nu}_1 (x) , \quad
-i {\de  W \ov \de A_{ 1\mu} (x)}=
\sqrt{-g_1} J^{\mu}_1 (x)  , \\
 i {\de W \ov \de g_{2\mu \nu} (x)}=
 \ha \sqrt{-g_2}  T^{\mu \nu}_2 (x) , \quad
i {\de  W \ov \de A_{ 2\mu} (x)}=
\sqrt{-g_2} J^{\mu}_2 (x)  \ .
 \label{defst2}
\end{gather}
The expectation values $T^{\mu \nu}, J^\mu$ of the stress tensor and
the $U(1)$ current  in {the} state $\rho_0$ in an external metric $g_{\mu \nu}$ and external background $A_\mu$ are obtained by
\be
T_{\mu \nu} = T^{\mu \nu}_1 \bigr|_{g, A} = T^{\mu \nu}_2   \bigr|_{g, A}  , \qquad  J^{\mu} = J^{\mu}_1  \bigr|_{g, A}   = J^{\mu}_2  \bigr|_{g, A}
\ee
where $\bigr|_{g, A}$ denotes setting  $g_{1\mu \nu} =  g_{2\mu \nu} =g_{\mu \nu}$ and $A_{1\mu} = A_{2\mu} =A_\mu$.


In the {\it absence} of any gravitational and $U(1)$ anomalies, $W[g_1, A_1; g_2, A_2]$ should be invariant under independent gauge transformations of $A_1, A_2$ and independent diffeomorphisms of $g_1 A_1$ and $g_2, A_2$, i.e.
\bega \label{in1}
W[g_1, A_1 + d \lam_1; g_2; A_2 + d \lam_2] = W[A_1, A_2]  \\
W[g_1, A_1; g_2, A_2] = W[g_1^{\xi_1}, A_1^{\xi_1}; g_2^{\xi_2}, A_2^{\xi_2}]
\label{in2}
\end{gather}
where $g^\xi, A^\xi$ denote diffeomorphisms of  $g, A$ generated by a vector field $\xi^\mu$.\footnote{$\lam_1, \lam_2$ and $\xi_{1,2}^\mu$ are all assumed to vanish at spatial and time infinities.} Equations~\eqref{in1}--\eqref{in2} in turn ensure that
\be \label{cons1}
\nab_{s \mu} J^\mu_s = 0, \qquad
\nab_{s \nu} T^{\nu}_{s\mu} =  F_{s\mu \nu} J^\nu_s , \quad s=1,2
\ee
where $\nab_1$ is the covariant derivative associated with $g_{1\mu \nu}$, and $F_{1\mu \nu}$ is the field strength of $A_{1\mu}$.
Similarly for quantities with subscript $2$.

For slowly varying sources, we can express the generating functional~\eqref{pager1} in terms of path integrals over
{\it slow} degrees of freedom of the system
\be \label{path1}
e^{W [ g_{1} , A_1; g_{2}, A_2]} = \int D \chi \, e^{{i \ov \hbar} I_{\rm EFT} [\chi]}
\ee
where $\chi$ collectively denotes slow variables of the system which in general also come in two copies.
The low energy effective action $I_{\rm EFT}$ depends on $\rho_0$ and external sources which we have suppressed, and is assumed to be local.

\begin{figure}[!h]
\begin{center}
\includegraphics[scale=0.9]{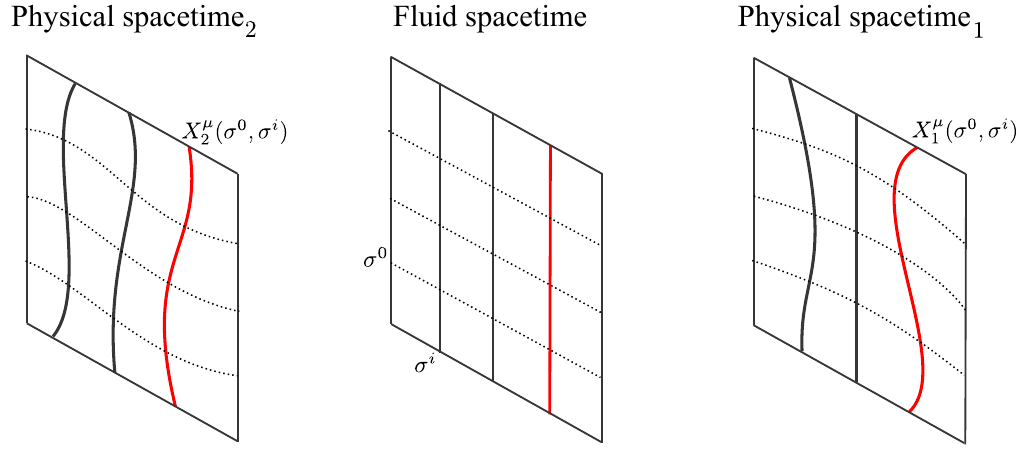}
\end{center}
\caption{$X^\mu_{1,2} (\sig^A)$ describe motion of  a continuum of fluid elements labelled by $\sig^A$ in two copies of physical spacetimes with coordinates $X^\mu_{1,2}$ respectively. $\sig^A = (\sig^0, \sig^i)$ are coordinates for a ``fluid spacetime'', where $\sig^i$ are interpreted as labels of each fluid element while $\sig^0$ is their ``internal'' time.
 The red straight
line in the fluid spacetime with constant $\sig^i$ is mapped by $X^\mu_{1,2} (\sig^0, \sig^i)$ to
physical spacetime trajectories (also in red) of the  corresponding fluid element.
 }
\label{fig:Lagrange}
\end{figure}

For $\rho_0$ describing a medium in local equilibrium, generically
the only  slow modes are those associated with conserved quantities~\eqref{cons1}, i.e.
hydrodynamical modes, with $I_{\rm EFT}$ the corresponding hydrodynamical action $I_{\rm hydro}$. We will limit ourselves to the generic situation.\footnote{The discussion can be readily generalized to systems such as near a critical point where one should also include the corresponding order parameter(s). See~\cite{GL,ping}.} The slow variables associated with the stress tensor can be chosen to be $X^\mu_{1,2} (\sig^A)$ which describe motions of  a continuum of fluid elements labelled by $\sig^A$ in two copies of physical spacetimes with coordinates $X^\mu_{1,2}$ respectively.
See Fig.~\ref{fig:Lagrange}. The slow variables associated with the $U(1)$ currents are  $\vp_{1,2} (\sig^A)$  which can be interpreted as $U(1)$ phase rotations  associated for each fluid elements.
It is also convenient to introduce an additional scalar field $\beta (\sig^A)$ which gives the local inverse temperature in fluid spacetime.\footnote{Note that there is only one temperature field rather than two copies.} $X^\mu_{1,2}$ and $\vp_{1,2}$ are
 the Stuckelberg fields for diffeomorphisms and gauge transformations~\eqref{in1}--\eqref{in2}, 
and we require the  hydrodynamical action $I_{\rm hydro}$ to be a local action of pullbacks of $g_{s \mu \nu}$ and $B_{s \mu} = A_{s \mu} + \p_\mu \vp_s$, $s=1,2$ to the fluid spacetime
 \bega \label{hdef1}
h_{sAB} (\sig)  =  {\p X^\mu_s \ov \p \sig^A} g_{s\mu \nu} (X_s (\sig)) {\p X_s^\nu \ov \p \sig^B} , \qquad
B_{sA} (\sig) = {\p X^\mu_s \ov \p \sig^A}  A_{s\mu} (X_s (\sig)) + \p_A \vp_s (\sig) \ .
\end{gather}
i.e.
  \be \label{act1}
I_{\rm hydro} = I_{\rm hydro} [h_1, B_1; h_2, B_2; \beta]  \ .
\ee

 By construction $h_{1,2}$ and $B_{1,2}$ are invariant under independent diffeomorphisms and gauge transformations
 of the two legs of the CTP contour
 ($s=1,2$):
\bega
 \label{1diffg}
g_{s \mu \nu}' (X_s') = {\p X_s^\lam \ov \p X_s'^\mu} {\p X_s^\rho \ov \p X_s'^\nu} g_{s \lam \rho} (X_s)   , \quad
A'_{s \mu} (X_s') = {\p X_s^\lam \ov \p X_s'^\mu} A_{s \lam} (X_s), \quad
X_s'^\mu (\sig)  = f_s^\mu (X_s (\sig))  \\
A_{s \mu}' = A_{s\mu} - \p_\mu \lam_s (X_s) , \qquad \vp_s' (\sig)= \vp_s (\sig) + \lam_s (X_s (\sig)) ,
\label{1ga}
\end{gather}
which along with~\eqref{act1}  immediately implies~\eqref{in1}--\eqref{in2}. 
Furthermore, the form of the action~\eqref{act1} implies that the equations of motion of $X^\mu_{1,2}$ and $\vp_{1,2}$ are equivalent to the conservations of
 the ``off-shell'' hydrodynamical stress tensors and currents defined as
\bega \label{defst}
{\de I_{\rm hydro} \ov \de g_{1\mu \nu} (x)} \equiv \ha \sqrt{-g_1} \hat T^{\mu \nu}_1 (x) ,\   \qquad
{\de I_{\rm hydro} \ov \de A_{1\mu} (x)}  \equiv \sqrt{-g_1} \hat J^{\mu}_1 (x) ,  \\
{\de I_{\rm hydro} \ov \de g_{2\mu \nu} (x)} \equiv - \ha \sqrt{-g_2} \hat T^{\mu \nu}_2 (x) ,\   \qquad
{\de I_{\rm hydro} \ov \de A_{2\mu} (x)}  \equiv -  \sqrt{-g_2} \hat J^{\mu}_2 (x) \ .
 \label{defst3}
\end{gather}

As defined the path integrals~\eqref{path1} apply to a general quantum system. 
At sufficiently high temperatures it is often enough to consider  the leading order in a small $\hbar$ expansion.
For this purpose we decompose
\bega \label{re1}
g_{1\mu\nu}=g_{\mu\nu}+\frac  \hbar2 g_{a\mu\nu} ,\quad g_{2\mu\nu} =g_{\mu\nu}-\frac \hbar2  g_{a\mu\nu}, \quad
A_{1\mu}=A_{\mu}+\frac \hbar 2 A_{a\mu} , \quad A_{2\mu}=A_{\mu}-\frac \hbar 2  A_{a\mu} \\
X_1^\mu =X^\mu+\frac \hbar2   X_a^\mu ,\quad  X_2^\mu =X^\mu-\frac \hbar 2  X_a^\mu ,\quad
\varphi_1 =\varphi+\frac \hbar 2   \varphi_a ,\quad  \varphi_2=\varphi-\frac \hbar 2 \varphi_a ,
\label{re2}
\end{gather}
and the action  $I_{\rm hydro}$ can be expanded in $\hbar$ as
\be
{1 \ov \hbar} I_{\rm hydro} = I_{\rm hydro}^{(0)} + \hbar I_{\rm hydro}^{(1)} + \cdots  \ .
\ee
In this limit the path integrals~\eqref{path1} survive and describe classical statistical averages. We will refer to variables with subscript $a$ as $a$-variables and those without as $r$-variables.
$r$-variables can be considered as describing physical quantities while $a$-variables correspond to noises. For example,
 $X^\mu (\sig^A)$ is  interpreted as mapping fluid spacetime into the physical spacetime (now only one copy) with $X_a^\mu$ interpreted as the corresponding position noises.

While the hydrodynamical action $I_{\rm hydro}$ is naturally formulated in the fluid spacetime $\sig^A$, one can also formulate it in physical spacetime by inverting $X^\mu (\sig^A)$, i.e. use $\sig^A (X)$ as dynamical variables and express all other variables accordingly as functions of $X^\mu$. In the physical spacetime formulation, the dynamical variables are then $\sig^A (x), \vp (x), \beta(x)$ and $ X^\mu_a (x), \vp_a (x)$, while the background fields are  $g_{\mu \nu} (x),  A_\mu(x), g_{a \mu \nu} (x), A_{a \mu} (x)$, where we have replaced $X^\mu$ by $x^\mu$ to emphasize they are now just coordinates for physical spacetime. The physical spacetime formulation has the advantage of being more physically intuitive and connects more directly with
the  traditional phenomenological approach.


\subsection{Formulation of $I_{\rm hydro}$ in physical spacetime}

We now list various symmetries and consistency requirements which $I_{\rm hydro}$  should satisfy
when formulated in the physical spacetime to leading order in the $\hbar$-expansion~\cite{CGL,CGL1,GL}. They can be separated into the following categories:

\ben

\item Spacetime diffeomorphisms and gauge transformations. In the absence of any gravitational and charged current anomalies, the action  $I_{\rm hydro}$ should be invariant under physical spacetime version of~\eqref{1diffg}--\eqref{1ga}.
Invariance under these transformations implies that $a$-fields (including both background and dynamical variables) must appear through the combinations
\bega
G_{a\mu\nu} (x)
 \equiv g_{a\mu \nu} + \sL_{X_a} g_{\mu \nu} = g_{a\mu \nu} + \nab_\mu X_{a \nu} + \nab_\nu X_{a \mu}
, \\
C_{a\mu} \equiv  A_{a\mu}(x )+ \p_\mu \vp_a (x) + \sL_{X_a} A_\mu
=  A_{a\mu}(x )+ \p_\mu \vp_a (x) +X_a^\nu \nab_\nu A_\mu + A_\nu \nab_\mu X_a^\nu
\label{pp2}
\end{gather}
while $A_\mu$ and $\vp$ must appear through
\be
B_\mu = A_\mu + \p_\mu \vp (x) \ .
\ee
The above variables are {the} physical spacetime version of~\eqref{hdef1}. 

\item Spatial and time diffeomorphisms in the fluid spacetime which {\it define} a fluid.  We require
the action $I_{\rm hydro}$ be invariant under
 \bega \label{sdiff}
 \sig^i \to  \sig'^i (\sig^i), \qquad \sig^0 \to \sig^0  \\
 \label{tdiff}
 \sig^0 \to \sig'^0= f (\sig^0, \sig^i), \qquad \sig^i \to \sig^i   \ .
 \end{gather}
Furthermore we require the action be invariant under the diagonal shift
\be\label{cshift}
\vp \to \vp - \lam (\sig^i (x^\mu)) , \qquad \vp_a \to \vp_a
\ee
where $\lam$ is a function of $\sig^i$ only. Invariance under~\eqref{cshift} defines a normal fluid. For a superfluid where the $U(1)$ symmetry is spontaneously broken  this symmetry should be dropped.  The symmetries~\eqref{sdiff}--\eqref{cshift} involve only dynamical variables, yet they should be viewed as ``global gauge symmetries,'' i.e. configurations related by such transformations are deemed physically equivalent.

Invariance under~\eqref{sdiff}--\eqref{cshift} implies that
the only invariant which can be constructed from $K_\mu^A \equiv \p_\mu \sig^A$ is the velocity field
$u^\mu$  defined by
\be \label{1vel}
u^\mu ={1 \ov b}  (K^{-1})^\mu_0 , \qquad b^2 = - g_{\mu \nu}  (K^{-1})^\mu_0  (K^{-1})^\nu_0
\ee
and $\De^{\mu \nu} = g^{\mu \nu} + u^\mu u^\nu$.
By definition $u^\mu u_\mu = -1$.
$B_\mu$ is not invariant under diagonal shift~\eqref{cshift} of $\vp$, but
\be
\mu \equiv u^\mu B_\mu, \qquad F_{\mu \nu} = \p_\mu B_\nu - \p_\nu B_{\mu} = \p_\mu A_\nu - \p_\nu A_\mu
\ee
are invariant. To summarize,  the only combinations of $r$-variables which can appear are
\be \label{ovar}
\beta (x) , \quad u^\mu, \quad \mu , \quad  F_{\mu \nu} , \quad \De^{\mu \nu} \ . 
\ee
It is often convenient to combine the first three variables further into
\be\label{uen}
\beta^\mu = \beta (x) u^\mu (x),  \qquad  \hat \mu (x) = \beta (x) \mu = \beta^\mu (x)  B_\mu
\ee
where  $\beta^\mu$ is now unconstrained. 

\item  Classical remnants of constraints from quantum unitarity of~\eqref{pager1}, 
\bega \label{keyp3}
  I^*_{\rm hydro} [\Lam_r, \Lam_a] = - I_{\rm hydro} [\Lam_r, - \Lam_a] \ , \\
 \label{keyp1}
  {\rm Im}  \, I_{\rm hydro}  \geq 0 \ , \\
  \label{keyp2}
 I_{\rm hydro}  [\Lam_r, \Lam_a =0 ] = 0  \ ,
\end{gather}
where $\Lam_{r,a}$ collectively denote all $r$- and $a$-variables including both dynamical and background fields.

\item  Discrete spacetime symmetries. If the microscopic system is invariant under charge conjugation $\sC$, parity $\sP$ or $\sC \sP$, such discrete symmetries should be imposed on $ I_{\rm hydro}$ and they can be imposed straightforwardly as usual.

\item We assume the microscopic Hamiltonian underlying the macroscopic many-body state $\rho_0$  is invariant under a discrete symmetry $\Th$ containing {\it time reversal}. $\Th$ can be time reversal $\sT$ itself, or any combinations of $\sC, \sP$ with $\sT$, such as $\sC\sP\sT$. $\Th$ can also be a combination of $\sT$ with some other internal discrete operations.
Unlike $\sC$ or $\sP$, $\Th$ by itself can {\it not} be imposed directly on  $I_{\rm hydro}$, since $\Th$ does not take the generating functional $W$ to itself, but to a time reversed generating functional $W_T$.\footnote{ This is quite intuitive as
 $I_{\rm hydro}$  contains dissipative terms, thus it cannot be invariant under $\Th$ alone.} The fact that the underlying Hamiltonian is invariant under $\Th$ nevertheless leads to important constraints on  $I_{\rm hydro}$ as we will discuss in the next item.

\item  We require $I_{\rm hydro}$ to be invariant under a $Z_2$ dynamical KMS symmetry
\be \label{bakv}
\tilde I_{\rm hydro} \equiv I_{\rm hydro} [\tilde \Lam_r, \tilde \Lam_a] = I_{\rm hydro} [\Lam_r,  \Lam_a]
\ee
where tilde denotes a $Z_2$ transformation which is a combination of  $\Th$ and
the Kubo-Martin-Schwinger (KMS) transformation.\footnote{As emphasized in~\cite{CGL}, when $\rho_0$ is given by a thermal density matrix, 
while neither $\Th$ nor the KMS operation takes the generating functional~\eqref{pager1} to itself, the generating functional $W$
is invariant under the combination of them 
i.e.
\be \label{wkms}
W [\tilde g_{\mu \nu}, \tilde A_\mu; \tilde g_{a \mu \nu} , \tilde A_{a \mu}] = W  [g_{\mu \nu}, A_\mu; g_{a \mu \nu} , A_{a \mu}] \ .
\ee
Accordingly in $I_{\rm hydro}$ one can {\it not} impose either $\Th$ or KMS separately, but should impose the combination of them~\eqref{bakv}.}

Equation~\eqref{bakv}  plays the dual role of imposing microscopic time-reversibility and {\it local equilibrium}.
It should be understood as a mathematical characterization of a state $\rho_0$  in local equilibrium. The prototype of such a state is the thermal density matrix in slowly varying external sources, but~\eqref{bakv} is more general, applicable also to  pure states. It was found in~\cite{CGL,CGL1,GL} that~\eqref{bakv} leads to Onsager relations, local first law, local second law, and local fluctuation-dissipation relations.

To leading order in $\hbar$, the tilde operation in~\eqref{bakv} can be written schematically as
\be \label{hyi}
\tilde \Lam_r = \Th  \Lam_r + O(\hbar), \qquad  \tilde \Lam_a =\Th  \Lam_a - i \Th  \Phi_r   + O(\hbar)
\ee
where $\Phi_r$ denotes certain combination of $r$-variables with total one derivative. More explicitly,
in~\eqref{hyi} we denoted $\Th$ transformation of a tensor $ G (x)$ as
\be \label{tg1}
\Th G (x) \equiv \eta_G G (\eta x) ,
\ee
where we have suppressed tensor indices for $G$, and $\eta_G$ should be understood as
a collection of phases ($\pm 1$) one for each component for $G$. Similarly for $\eta x$. For example, for $\Th = \sT$ and $G = A_\mu$
\be
\eta_A A_\mu = (A_0, - A_i) , \quad \eta x^\mu = (- x^0, x^i),  
\ee
while for $\Th = \sC \sP \sT$
\be
\eta_A A_\mu = (-A_0, - A_i) , \qquad \eta x^\mu = (- x^0, -x^i) \ .  
\ee
Since $\Th$ contains $\sT$
\be
\Th i = - i \Th  \ .
\ee
The second set of equations in~\eqref{hyi} for $a$-variables can be written explicitly as
\bega \label{si2}
\Th \tilde \vp_a (x) =  \vp_a (x) +
 i \beta^\mu \p_\mu \vp  (x), \\
\Th \tilde G_{a\mu\nu} (x) =  G_{a\mu \nu}  (x) + i \sL_{\beta^\mu} g_{\mu \nu} (x)
=G_{a\mu \nu}  (x) + i \le(\nab_{\mu} \beta_{\nu} + \nab_{\nu} \beta_{\mu} \ri) ,  \\
\Th \tilde C_{a\mu} (x) =  C_{a \mu} (x) + i \sL_{ \beta^\mu} B_\mu  (x)
= C_{a \mu} (x) + i \le(\nab_\mu \hmu - \beta^\nu F_{\mu \nu}  \ri)
\ .
\label{si22}
\end{gather}
The explicit transformations for $\Th= \sT, \sP \sT, \sC \sP \sT$ for various tensors  are given in Appendix~\ref{app:cpt}.


\een

It is straightforward to write down the most general $ I_{\rm hydro} = \int d^d x \, \sqrt{-g} \, \sL$ consistent with the above prescriptions.
We can expand the corresponding Lagrangian density $\sL$ in terms of the number of
$a$-variables and derivatives. The  first few terms in the $a$-field expansion can be written schematically  as 
\be \label{acttja}
\mathcal L = \ha \hat T^{\mu M}G_{a\mu M}+ {i \ov 4} W^{\mu\nu,MN} G_{a\mu M}G_{a\nu N}+ {1 \ov 8} Y^{\mu\nu\rho,MNP}G_{a\mu M}G_{a\nu N}G_{a\rho P}+\cdots \ ,
\ee
where we have introduced notation
\be \label{n0}
G_{a\mu M}=(G_{a\mu\nu},2 C_{a\mu}), \quad
\hat T^{\mu M}=(\hat T^{\mu\nu}, \hat J^{\mu}), \quad M = (\mu, d), \quad G_{a \mu d} = 2 C_{a \mu} \
\ee
and $\hat T^{\mu M}, W^{\mu\nu,MN}, \cdots$ are covariant tensors constructed out of $r$-variables
 $\{\beta^\mu, \hat \mu, F_{\mu \nu}, \De^{\mu \nu}\}$ and covariant derivatives on $G_{a \mu M}$.
Given that $G_{a \mu \nu} = g_{a \mu \nu} + \cdots$ and  $C_{a \mu} = A_{a \mu} + \cdots $,  we identify
$\hat T^{\mu \nu}$ and $\hat J^\mu$  as the ``off-shell'' hydrodynamic stress tensor and $U(1)$ current, and {the} equations of motion of $X^\mu_a, \vp_a$ give the standard hydrodynamic equations.

If we introduce  $n$ as the sum of the number of $a$-fields and the number of derivatives
in a term, then since $\Phi_{r }$ in~\eqref{hyi} contains one derivative, the dynamical KMS transformation~\eqref{bakv}   preserves $n$, which implies
that terms in the action which have the same value of $n$ transform separately among themselves.
We can thus write the action as
\be \label{altex}
 \sL =  \sum_{n=1}^\infty  \sL_n = \sL_1 + \sL_2 + \sL_3 + \cdots
\ee
where $\sL_n$ contains all terms with given $n$. They are separately invariant under~\eqref{bakv}.
$\sL_1$ contains only zeroth derivative term in $ \hat T^{\mu M}$ while $\sL_2$ contains first derivative terms
in $\hat T^{\mu M}$ and zeroth derivative terms in $W^{\mu\nu,MN}$. The explicit expressions for~\eqref{acttja} to order $\sL_2$ for a parity-preserving fluid are given in~\cite{CGL1}.\footnote{They are given to order $\sL_3$ for conformal fluids.}

We now give a brief review of the derivation of the entropy current, whose details are given in \cite{GL}. Dynamical KMS invariance (\ref{bakv}) implies that
\be\label{kmsi0} \tilde{\mathcal L}=\mathcal L+\nabla_\mu V^\mu,\qquad V^\mu=V_0^\mu+V_1^\mu+\cdots\ee
where $\tilde{\mathcal L}=\mathcal L[\Theta\tilde\Lambda_a,\Theta\tilde\Lambda_r]$, and
$V_k^\mu$ contains $k$ factors of $a$-fields.
%
The entropy current can then be defined as
\be \label{nmj}
 S^\mu = V^\mu_0-\hat V_1^\mu-\hat T^{\mu\nu} \beta_\nu-\hat\mu \hat J^\mu \ ,
 \ee
where $\hat V_1^\mu$ is $V_1^\mu$ with $\Lambda_a$ replaced by the corresponding $\Phi_r$ as introduced in (\ref{hyi}).
It can be shown upon using equations of motion
\be
\nabla_\mu S^\mu = R \geq 0
\ee
 where $R$ is a local non-negative expression.


\section{Effective action for parity-violating systems in (3+1)-dimension} \label{sec:31}

In this section we apply the formalism reviewed in the previous section to four-dimensional systems which break parity, including those with a local $U(1)$ anomaly
\be \label{j0}
\nab_\mu \hat J^\mu = {c \ov 4} \hbar \ep^{\mu \nu \lam \rho} F_{\mu \nu} F_{\lam \rho}   \
\ee
where 
constant $c$ depends on specific systems.
$\ep^{\mu \nu \lam \rho}$ is the fully antisymmetric tensor with $\ep^{0123} = {1 \ov \sqrt{-g}}$. 
In~\eqref{j0} we have made manifest $\hbar$-dependence so as to be clear about the order in $\hbar$-expansion at which the corresponding anomalous transports appear in the hydrodynamical action.
We assume that
the system does {\it not} have any {\it local} mixed gravitational anomalies. We will see that the system can nevertheless
possess {\it global} gravitational anomalies which are closely connected to certain novel transports.


\subsection{Generating functional}

 From~\eqref{j0}, under independent local transformations of $A_{1,2}$, equation~\eqref{in1} should be replaced by\footnote{We emphasize that here we consider only small gauge transformations and diffeomorphisms, i.e. those vanish at spatial and time infinities and smoothly connected to the identity.}
\be \label{aneq}
- i  W [g_1, A_1 - d \lam_1  ;  g_2 , A_2 - d \lam_2,]  =
-i W [A_1, g_1; A_2, g_2 ] + c  \int \le(\lam_1 F_1 \wedge F_1 -  \lam_2  F_2 \wedge F_2 \ri)
\ee
while~\eqref{in2} remains.  Note that $F \equiv \ha F_{\mu \nu} dx^\mu \wedge dx^\nu = dA$, and the second term on the right hand side is independent of metrics.
Indeed, from~\eqref{aneq} the consistent currents introduced in~\eqref{defst1}--\eqref{defst2} now satisfy\footnote{Note that
when restoring $\hbar$, there should be a $\hbar$ factor on the left hand side of various equations in~\eqref{defst1}--\eqref{defst2}.}
\be \label{j1}
\nab_\mu J^\mu = {c \hbar \ov 4 } \ep^{\mu \nu \lam \rho} F_{\mu \nu} F_{\lam \rho}
\ee
and from diffeomorphism invariance of $W$ we also have
\be \label{ste1}
\nab_{\nu} T^{\nu}{_\mu} =  F_{\mu \nu} J^\nu - A_\mu \nab_\nu J^\nu =
F_{\mu \nu} J^\nu -{c \ov 4} \hbar A_\mu  \epsilon^{\alpha\beta\gamma\delta}F_{\alpha\beta}F_{\gamma\delta} \ .
\ee
In~\eqref{j1}--\eqref{ste1} we have suppressed indices $1,2$. It should be understood there are two copies of them and
so are~\eqref{10c}--\eqref{12c} below.
Defining the covariant current as
\be  \label{10c}
\sJ^\mu = J^\mu + {c \hbar } \epsilon^{\mu \nu \rho \lam} A_\nu F_{ \rho \lam}
\ee
we can write equations~\eqref{j1} and~\eqref{ste1} as
\begin{gather}
\label{11c}
 \nab_{\mu} \sJ^\mu ={3  c\ov 4} \hbar \epsilon^{\alpha\beta\gamma\delta} F_{ \alpha\beta}F_{ \gamma\delta} , \\
\label{12c}
\nab_{\nu} T^{\nu}{_\mu} =  F_{\mu \nu} \sJ^\nu \ .
\end{gather}
Note that the equation for $T^{\mu \nu}$ must be expressible in terms of covariant current $\sJ^\mu$ as $T^{\mu \nu}$ should be gauge invariant (the last term in~\eqref{aneq} is independent of the metric).  To leading order in $\hbar$-expansion, the anomalous piece in~\eqref{aneq} becomes (see~\eqref{re1}--\eqref{re2} and $\lam_a = \lam_1 - \lam_2$)
\be
c \int \le(\lam_1 F_1 \wedge F_1 -  \lam_2  F_2 \wedge F_2 \ri)  =
 c \hbar \int \le(\lam_a F \wedge F + 2 \lam F \wedge F_a \ri)  + O(\hbar^2) \ .
\label{wvar}
 \ee

\subsection{Parity odd action}

We now construct the hydrodynamic action for a parity-violating system with a local $U(1)$ anomaly.
We can write the action as
 \be \label{iac}
I_{\rm hydro} =  I_{\rm even} + I_{\rm odd}  
\ee
where $I_{\rm even}$ and $I_{\rm odd}$ are  parity even and odd parts respectively.
$I_{\rm odd}$ can be further decomposed as
\be \label{iac1}
 I_{\rm odd} = I_{\rm o, inv} + I_{\rm anom}
 \ee
where $I_{\rm anom} $ is responsible for generating the anomalous term in~\eqref{aneq}, and $I_{\rm o, inv}$
is invariant under gauge transformations. Given that $I_{\rm even} $ is invariant under gauge transformations we can also write
\be \label{jke}
I_{\rm hydro} =  I_{\rm inv}  + I_{\rm anom} , \qquad I_{\rm inv} =  I_{\rm o, inv} +  I_{\rm even} \ .
\ee
Note that $I_{\rm inv}$ should depend on $\vp_{1,2}$ only through $B_{1,2}$ introduced in~\eqref{hdef1}, while $I_{\rm anom}$ does not have to.

 Since neither the diagonal shift~\eqref{cshift} nor the dynamical KMS transformations~\eqref{hyi} mix parity even and odd parts,
$I_{\rm even}$ and $I_{\rm odd}$ can be treated independently. $I_{\rm even}$ was discussed in detail in~\cite{CGL,CGL1}. Here we focus on 
\be
I_{\rm odd} = \int d^4 x \, \sqrt{-g} \, \sL_{\rm odd} \ .
\ee
and will construct $\sL_{\rm odd}$ to order
$\sL_2$ as defined in~\eqref{altex}.



Let us first look at $I_{\rm anom}$. To match with the anomalous term in~\eqref{aneq}, we take the anomalous action as (written in fluid spacetime)
\begin{align} 
{1 \ov  \hbar } I_{\rm anom } 
& =   c\int 
\le[\vp_1 F_{1} (X_1) \wedge F_{1} (X_1) - \vp_2 F_{2} (X_2) \wedge  F_{2} (X_2) \ri]
\label{anom}
\end{align}
where $X_{1,2}^\mu$ are functions of $\sig^A$, $F_{1AB}$ is the pull-back of $F_{1 \mu \nu}$.
Note that under  gauge transformations~\eqref{1ga}
we 
precisely recover~\eqref{aneq} from~\eqref{anom}. To see this, for two terms in~\eqref{anom} one changes the integration variables to $X_1$ and $X_2$ respectively, which then become dummy variables.

Given~\eqref{anom}  and that $I_{\rm inv}$ depends only on $B_{1,2}$, the equations of motion of $\vp_s$ and $X_s^\mu$ lead to
\begin{gather}
\label{1eom11}
 \nab_{\mu} \hat J^\mu = {c \ov 4} \hbar \epsilon^{\alpha\beta\gamma\delta}F_{ \alpha\beta}F_{ \gamma\delta} , \\
\label{1eom12}
\nab_{\nu} \hat T^{\nu}{_\mu} =  F_{\mu \nu}\hat J^\nu -{c \ov 4}  \hbar A_\mu \epsilon^{\alpha\beta\gamma\delta} F_{\alpha\beta}F_{\gamma\delta} \ ,
\end{gather}
where the off-shell stress tensors and consistent currents are defined in~\eqref{defst}--\eqref{defst3}. Again we have suppressed $s=1,2$
and each equation should be understood to have two copies.  Defining the covariant off-shell currents as
\be \label{cov1}
\hat \sJ^\mu = \hat J^\mu + { c \hbar } \epsilon^{\mu \nu \rho \lam} A_\nu F_{ \rho \lam}
= \hat J_{\rm inv}^\mu + { c \hbar} \epsilon^{\mu \nu \rho \lam} B_\nu F_{ \rho \lam}
\ee
where $\hat J_{\rm inv}^\mu$ is defined as the off-shell currents corresponding to $I_{\rm inv}$,
we then have
\begin{gather}
\label{1eom11c}
 \nab_{\mu} \hat \sJ^\mu ={3  c \ov 4} \hbar \epsilon^{\alpha\beta\gamma\delta} F_{ \alpha\beta}F_{ \gamma\delta} , \\
\label{1eom12c}
\nab_{\nu} \hat T^{\nu}{_\mu} =  F_{\mu \nu}\hat \sJ^\nu \ .
\end{gather}
The last equality of~\eqref{cov1} makes it manifest that $\hat \sJ^\mu$ is invariant under gauge transformations.

Expanding in small $\hbar$ and rewriting the resulting expressions in physical spacetime we find that~\eqref{anom} becomes 
\be \label{ianm}
{1 \ov \hbar} I_{\rm anom} = c\hbar  \int d^4 x \,\le( \vp_a F \wedge F +2 \vp F \wedge \sF_a  \ri)
\ee
where
\be \label{fdef}
\sF_a = d C_a = F_a + \sL_{X_a} F  
\ee
and $C_{a \mu}$ was defined in~\eqref{pp2}. Note that ${1 \ov \hbar}  I_{\rm anom}$ is of order $O(\hbar)$.
Under a diagonal shift~\eqref{cshift}, equation~\eqref{ianm} transforms
as
\be \label{var1}
{1 \ov \hbar}  \de I_{\rm anom} 
= 2c  \hbar \int  \lam (\sig^i) F \wedge \sF  \ .
\ee
In order for the full odd action~\eqref{iac1} to be invariant under~\eqref{cshift},  $I_{\rm o, inv}$  should also not be invariant
and its variation should precisely cancel~\eqref{var1}. 

At linear order in $a$-fields (order $O(a)$) we can write
\be \label{oddac}
{1 \ov \hbar}  \sL_{\rm o, inv} =  \ha \hat T^{\mu \nu}_o G_{a \mu \nu}  + \hat J^\mu_o C_{a \mu}  \
\ee
and the terms on the right hand side may be further expanded in $\hbar$ and derivatives.

Let us first consider $\hat T^{\mu \nu}_o$ which as usual can be decomposed as
\be
\hat T^{\mu \nu}_o = \vep_o u^\mu u^\nu + p_o \De^{\mu \nu} + 2 u^{(\mu} q^{\nu)}_o + \Sig_o^{\mu \nu} , \qquad
\ee
where $q^{\nu}_o$ and  $\Sig_o^{\mu \nu}$ are transverse to $u^\mu$. Since  terms proportional to $G_{a \mu \nu}$ will never generate a term of the form~\eqref{var1} under~\eqref{cshift}, $T^{\mu \nu}_o$ should be diagonal shift invariant by itself.  At zeroth derivative order there is no such term.  At first derivative order the only non-vanishing quantity 
is $q^{\nu}_o$ which can
 be written as
\be \label{owp0}
q^{\mu}_o = g_1 \om^\mu + g_2 \sB^\mu
\ee
where
\be \label{ombd}
\om^\mu \equiv \ha \ep^{\mu \nu \lam \rho} u_\nu \le(\nab_\lam u_\rho - \nab_\rho u_\lam \ri) , \qquad
\sB^\mu \equiv  \ha \ep^{\mu \nu \lam \rho} u_\nu  F_{\lam \rho} 
\ee
and $g_1,g_2$ are some functions of $\beta (x)$ and $\hmu (x)$.

$\hat J^\mu_o$ can be written as
\be \label{owp}
\hat J^\mu_o = - c  \hbar \ep^{\mu \nu \lam \rho}   F_{ \nu \lam} B_\rho  + \hat j^\mu_o
\ee
where the variation of the first term under diagonal shift cancels~\eqref{var1} and thus $\hat j^\mu_o$
should be invariant. From~\eqref{cov1} we thus find that
\be
\hat \sJ^\mu_{\rm odd} =  \hat j^\mu_o \ .
\ee

As discussed above to first derivative order since there is no diagonal shift invariant scalar term
$\hat j^\mu_o$ should then be transverse and can be written as
\be \label{owp1}
\hat j^\mu_o =
  h_1 \om^\mu + h_2 \sB^\mu
\ee
where $h_1,h_2$ are some functions of $\beta (x)$ and $\hmu (x)$.

Now let us consider quadratic terms in $a$-fields (order $O(a^2)$) 
to zeroth order in derivative, which should have the form
\be
 {i \ov 4} W^{\mu\nu,MN}_o G_{a\mu M}G_{a\nu N}
\ee
where $W^{\mu\nu,MN}_o $ is parity odd and is diagonal shift invariant. Such a term does not exist at zero derivative order
so we conclude there are no new parity-odd terms at order $O(a^2)$. 


Collecting the above expressions,  $\sL_{\rm odd}$ can be written as
\ie
{1 \ov \hbar} \sL_{\rm odd} & =  {c \ov 4} \hbar \ep^{\mu \nu \lam \rho} \le( \vp_a F_{\mu \nu}  F_{\lam \rho} +
2 \vp F_{\mu \nu} \sF_{a \lam \rho} \ri)
+  u^{(\mu} q^{\nu)}_o  G_{a \mu \nu}  +\le(\hat j^\mu_o - c  \hbar \ep^{\mu \nu \lam \rho}   F_{ \nu \lam} B_\rho \ri) C_{a \mu} \cr
& = u^{(\mu} q^{\nu)}_o  G_{a \mu \nu} + \le(\hat j^\mu_o - c  \hbar \ep^{\mu \nu \lam \rho}   F_{ \nu \lam} A_\rho \ri) C_{a \mu}  +
 {c \ov 4} \hbar \ep^{\mu \nu \lam \rho}  \vp_a F_{\mu \nu}  F_{\lam \rho} 
\label{finac}
\fe
where $\sF_{a \lam \rho}$ is defined by~\eqref{fdef}, and $q^\mu_o$, $\hat j^\mu_o$ are given respectively by~\eqref{owp0},~\eqref{owp1}. Using field redefinitions one can write $\sL_{\rm odd}$ in the Laudau frame (see Sec. VI of~\cite{CGL1} for details)
\be \label{lanac}
\sL_{\rm odd} =  \le(\ell^\mu_o - c  \hbar \ep^{\mu \nu \lam \rho}   F_{ \nu \lam} A_\rho \ri) C_{a \mu}  +
 {c \ov 4} \hbar \ep^{\mu \nu \lam \rho}  \vp_a F_{\mu \nu}  F_{\lam \rho}
\ee
with $\ell^\mu_o$ given by
\be \label{oii}
\ell^\mu_o \equiv \hat j^\mu_o - {n_0 \ov \ep_0 + p_0}  q^\mu_o  \
\ee
where $\ep_0, p_0, n_0$ are respectively zeroth order energy, pressure and charge densities.

\subsection{Dynamical KMS condition}

We now impose the dynamical KMS condition~\eqref{bakv} on the parity-odd action~\eqref{iac1}. We will consider respectively $\Th = \sP\sT, \sT, \sC \sP \sT$ and will see that they lead to very different results.


Due to the presence of $\hbar$ on the right hand side of~\eqref{j1}, ${1 \ov \hbar} I_{\rm anom}$ is of order $O(\hbar)$.
In ${1 \ov \hbar} I_{\rm o, inv}$ the first term in~\eqref{owp}
is $O(\hbar)$ while $g_1, g_2, h_1, h_2$ are undetermined at the moment. We will later argue that they should also be $O(\hbar)$.
Thus in our discussion below it is enough to consider the leading order terms in dynamical KMS transformations~\eqref{hyi}.\footnote{In fact one can check that the structure of $O(\hbar)$ corrections in~\eqref{hyi} are such that even if $g_1, g_2, h_1, h_2$ are of order $\hbar^0$, at first derivative order for $O(a)$ terms the leading terms in~\eqref{hyi} are adequate.}



\subsubsection{$\Th = \sP \sT$}

We find in this case
\begin{align}
\frac{1}{\hbar}\tilde{\mathcal{L}}_{odd}-\frac{1}{\hbar}\mathcal{L}_{odd}&=-2\frac{\mathcal{L}_{odd}}{\hbar}-u^\mu q_o^\nu\nabla_\mu\beta_\nu\\&-(\hat{j}_o^\mu-c\hbar\epsilon^{\mu\nu\lambda\rho}F_{\nu\lambda}A_\rho)(\partial_\mu\hat{\mu}+\beta^\alpha F_{\alpha\mu})-\frac{c\hbar}{4}\epsilon^{\mu\nu\rho\sigma}F_{\mu\nu}F_{\rho\sigma}\beta^\alpha\partial_\alpha\phi
\end{align}
KMS invariance at $\mathcal{O}(a)$ then requires
\be
g_1=h_1=g_2=h_2=c=0 \ .
\ee


\subsubsection{$\Th = \sT$}

From~\eqref{ianm} we find that under dynamical KMS transformation, the anomalous action becomes ({see Appendix~\ref{app:a} for useful formulae})
\bea
{\frac{\tilde I_{\rm anom}}{\hbar}} &=&    c \hbar \int \le( (\vp_a  + \sL_{i \beta} \vp)  F \wedge F +2 \vp F \wedge (\sF_a + \sL_{i \beta} F) \ri) \cr
& = &   {\frac{I_{\rm anom}}{\hbar}} + 2 i c \hbar  \int  \, d \le(\vp (\beta \cdot F) \wedge F \ri)
\label{jio}
\eea
where $\beta \cdot F \equiv \beta^\mu F_{\mu \nu} dx^\nu$.
For dynamical KMS transformation of $I_{\rm o, inv}$, note that
\be
\tilde q_o^\mu (-x^0, x^i) = (q_o^0 , - q_o^i)  (x), \qquad \tilde{\hat{J_o}}^\mu (-x^0, x^i) = (\hat J_o^0, - \hat  J_o^i) (x) \ .
\ee
We then find that
\be \label{tiod}
{\frac{\tilde I_{\rm o,inv}}{\hbar}} =  {\frac{I_{\rm o, inv}}{\hbar}} + i \int d^4 x \, \le(\ha \hat T^{\mu \nu}_o \sL_{\beta} g_{\mu \nu}  + \hat J^\mu_o \sL_{ \beta} B_{\mu} \ri) \ .
\ee
For $I_{\rm odd}$ to be invariant, we need the second term of~\eqref{tiod} to be a total derivative. More explicitly, using~\eqref{owp0}--\eqref{owp1}, we find after some algebraic manipulations  ({see Appendix~\ref{app:a} for useful formulae})
\bea
&& i \int d^4 x \, \le(\ha \hat T^{\mu \nu}_o \sL_{\beta} g_{\mu \nu}  + \hat J^\mu_o \sL_{ \beta} B_{\mu} \ri) \cr
&= & i \int \le[\le({\beta h_2 \ov 2}+ 3 c \hbar  \hmu \ri)  F \wedge F +  (h_2 d \hmu - g_2 d \beta) \wedge u \wedge F +  (h_1 d \hmu  - g_1 d \beta) \wedge u \wedge du \ri. \cr
&& \le. \qquad + \;  h_1 (\beta \cdot F)
\wedge u \wedge du +  g_2 (\beta \cdot \om) \wedge u \wedge F + {g_1 \beta \ov 2} du \wedge du  -2 c \hbar d (\hmu B \wedge F ) \ri]
\label{tot}
 \eea
 where $u \equiv u_\mu dx^\mu$.
For the above expression to be a total derivative we find that $h_1, h_2, g_1, g_2$ must arise from derivatives of two functions $H_1, H_2$ and satisfy the following relations
\bega \label{tot1}
h_2 =  {-6 c \hbar \hmu + 2 a_1\ov \beta}, \quad h_1 = g_2 , \quad \p_\hmu H_2 = h_2, \quad \p_\beta H_2 = - g_2 , \quad
\p_\hmu H_1 = h_1, \\
 \p_\beta H_1 = - g_1 , \quad 2 H_1  = g_1 \beta , \quad g_2 \beta = H_2
 \label{tot3}
\end{gather}
where $a_1$ is a constant. Note that one could add a constant to the right hand side of equation $2 H_1 = g_1 \beta$, but that constant can be absorbed in the definition of $H_1$. Similarly with equation $g_2 \beta = H_2$.  With~\eqref{tot1}--\eqref{tot3},
\bega \label{tot2}
 i \int d^4 x \, \le(\ha \hat T^{\mu \nu}_o \sL_{\beta} g_{\mu \nu}  + \hat J^\mu_o \sL_{ \beta} B_{\mu} \ri)
=  i \int  d Q, \\
Q =  -2 c \hbar \hmu B \wedge F + a_1 A \wedge F
+ H_2 u \wedge F +  H_1 u \wedge du \ .
\label{jio1}
\end{gather}
Note that $Q$ is defined only up to a closed three-form as such an addition will not change~\eqref{tot2}.

The most general solutions to~\eqref{tot1}--\eqref{tot3} can be written as
\bega \label{HH}
H_2 =  {- 3 c \hbar \hmu^2 + 2 a_1 \hmu + a_2 \ov \beta}  , \qquad H_1 = {- c \hbar \hmu^3 + a_1 \hmu^2 + a_2 \hmu + a_3 \ov \beta^2} \\
 \label{hh}
h_1 = {-3 c \hbar \hmu^2 + 2 a_1 \hmu + a_2  \ov \beta^2} , \quad h_2 =  {- 6c \hbar \hmu + 2 a_1 \ov \beta} ,\\
g_1 =  {-2 c \hbar \hmu^3 + 2 a_1 \hmu^2 +  2 a_2 \hmu + 2 a_3 \ov \beta^3} , \quad g_2 =   {- 3c \hbar \hmu^2 + 2 a_1 \hmu + a_2 \ov \beta^2} \
\label{gg}
\end{gather}
where  $a_1,a_2,a_3$ are constants. Thus to first derivative order $I_{\rm odd}$ is fully determined up to three  constants.

\subsubsection{$\Th = \sC \sP \sT$}

The analysis for $\Th = \sC \sP \sT$ is very similar. Note that
\be
\tilde F_{\mu \nu} (-x) = F_{\mu \nu} , \qquad \tilde \sF_{a \mu \nu} (- x) =
\sF_{a \mu \nu} (x) + i  \sL_{ \beta} F_{\mu \nu} (x)
\ee
and equation~\eqref{jio} again applies.
For $I_{\rm o, inv}$, we now have
\bega
\tilde q_o^\mu (- x) =  - g_1 (-\hmu, \beta) \om^\mu (x) + g_2  (-\hmu, \beta) \sB^\mu (x) \\
 \tilde J_o^\mu (-x) =   c \hbar \ep^{\mu \nu \lam \rho}   F_{ \nu \lam} B_\rho  - h_1 (-\hmu, \beta) \om^\mu (x)+ h_2 (-\hmu, \beta) \sB^\mu  (x) \ .
\end{gather}
and the dynamical KMS condition at $O(a)$ level requires
\be \label{pou}
g_1 (-\hmu) = - g_1 (\hmu), \quad  g_2 (-\hmu) = g_2 (\hmu), \quad h_1 (-\hmu) =  h_1 (\hmu), \quad
 h_2 (-\hmu) = - h_2 (\hmu)  \ .
\ee
The analysis for $O(a^0)$ terms is the same as before and~\eqref{tot1}--\eqref{jio1} apply.  Imposing~\eqref{pou} on the solutions~\eqref{HH}--\eqref{gg} we find that $a_1 = a_3 =0$, and thus
\bega \label{gh1}
h_1 = {-3 c\hbar  \hmu^2  + a_2  \ov \beta^2} , \quad h_2 =  - {6c \hbar \hmu\ov \beta} ,\quad
g_1 =  {-2 c \hbar \hmu^3  +  2 a_2 \hmu  \ov \beta^3} , \quad g_2 =   {- 3 c \hbar \hmu^2  + a_2 \ov \beta^2} ,  \\
H_2 =  {- 3c \hbar \hmu^2  + a_2 \ov \beta}  , \qquad H_1 = {- c \hbar \hmu^3  + a_2 \hmu  \ov \beta^2}  \ .
\end{gather}
Thus for a macroscopic system whose underlying Hamiltonian is invariant under $\sC \sP \sT$  to first derivative order $I_{\rm odd}$ is fully determined up to a single constant.

\subsection{Explicit  expressions for $q_o^\mu$ and $\hat j_o^\mu$}

We can now write down the explicit expressions for $q_o^\mu$ and $\hat j_o^\mu$ to be used in~\eqref{finac} or~\eqref{lanac}.
It is enough to do it for $\Th = \sT$. The expressions for $\Th = \sC \sP \sT$ can be obtained by setting $a_1 = a_3 =0$, while those for $\Th = \sP \sT$ can be obtained by setting $a_1 = a_2 = a_3$ to zero.

From~\eqref{hh}--\eqref{gg} we find that
\bega \label{qo0}
q^\mu_o = - c \hbar {\hmu^2  \ov \beta^2} \le({2 \hmu \ov \beta}  \om^\mu + 3 \sB^\mu \ri)
+  {2 a_1 \hmu \ov \beta^2} \le({\hmu \ov \beta}  \om^\mu + \sB^\mu \ri) + {a_2 \ov \beta^2}
\le({2 \hmu \ov \beta}  \om^\mu + \sB^\mu \ri) + {2 a_3 \ov \beta^3} \om^\mu  \\
\hat j_o^\mu = -3c \hbar {\hmu \ov \beta} \le({\hmu \ov \beta}  \om^\mu + 2 \sB^\mu \ri)
+  {2 a_1 \ov \beta} \le({\hmu \ov \beta}  \om^\mu + \sB^\mu \ri) + {a_2 \ov \beta^2} \om^\mu \ .
\label{qo1}
\end{gather}
The frame independent quantity $\ell^\mu_o$~\eqref{oii}
is then given by
\ie \label{po1}
\ell^\mu_o & = -3 c \hbar {\hmu \ov \beta} \le[\le(1 - {2 \al \ov 3}\ri) {\hmu \om^\mu \ov \beta} + (2 - \al) \sB^\mu\ri]
+ {2 a_1 \ov \beta} (1 - \al) \le({\hmu \ov \beta}  \om^\mu + \sB^\mu \ri) \cr
&\qquad + {a_2 \ov \beta^2} \le[(1 - 2 \al) \om^\mu  - {n_0 \ov  \ep_0 + p_0} \sB^\mu \ri] - {2 a_3 \ov \beta^3} {n_0 \ov  \ep_0 + p_0}  \om^\mu
\\
& = \xi_\om \om^\mu + \xi_B \sB^\mu
\fe
where we have introduced
\be
\al \equiv {n_0 \hmu \ov \beta (\ep_0 + p_0)}
\ee
and
\ie \label{po2}
\xi_\om & = -3 c \hbar {\hmu^2 \ov \beta^2} \le(1 - {2 \al \ov 3}\ri) + {2 a_1 \hmu  \ov \beta^2} (1 - \al)
+ {a_2 \ov \beta^2} (1 - 2 \al) - {2 a_3 \ov \beta^3} {n_0 \ov  \ep_0 + p_0}  , \\
\xi_B & = -3 c \hbar {\hmu \ov \beta}  (2 - \al) + {2 a_1 \ov \beta} (1 - \al)  - {a_2 \ov \beta^2}   {n_0 \ov  \ep_0 + p_0} \ .
\fe

Equations~\eqref{po1}--\eqref{po2} reproduce previous results in the literature obtained from entropy current~\cite{Son:2009tf},~\cite{Neiman:2010} and
equilibrium partition function~\cite{Banerjee:2012iz}, confirming that these methods indeed give the complete answer for the current problem. However, those methods did not pinpoint the exact discrete symmetry a system should have
for~\eqref{po1}--\eqref{po2}. Ref.~\cite{Banerjee:2012iz} did point out for $\sC \sP \sT$ invariant theories one should set $a_1 = a_3 =0$.

We presented our results in terms of $\om^\mu, \sB^\mu$ which were defined in~\eqref{ombd} from respective ``field strengths'' of $u_\mu$ and $B_\mu$. But note that $u^\mu B_\mu \neq 0$.
We now present~\eqref{qo0}--\eqref{po1} in a slightly different basis which makes their expressions a bit more transparent. Introduce
\be \label{bewb}
b_\mu = \De_\mu^\nu B_\nu = B_\mu + \hmu v_\mu, \qquad v_\mu = {u_\mu \ov \beta} , \qquad b_\mu v^\mu = 0
\ee
and 
\be
\fw^\mu =  \ha \ep^{\mu \nu \lam \rho} v_\nu \le(\nab_\lam v_\rho - \nab_\rho v_\lam \ri) , \qquad
\fB^\mu=  \ha \ep^{\mu \nu \lam \rho} v_\nu  \le(\nab_\lam b_\rho - \nab_\rho b_\lam \ri)   
 \ .
\ee
Note that $v^\mu b_\mu =0$, and
\be
\om^\mu = \beta^2 \fw^\mu, \qquad \sB^\mu = \beta \le(\fB^\mu - \hmu \fw^\mu \ri) \ .
\ee
Then equations~\eqref{qo0}--\eqref{qo1} and~\eqref{po1} can be rewritten as
\bega \label{m0}
q^{\mu}_o = {c \hbar \hmu^2 \ov \beta} \le( \hmu \fw^\mu - 3 \fB^\mu \ri) + {2 a_1 \hmu \ov \beta} \fB^\mu
+ {a_2\ov \beta}  \le( \hmu \fw^\mu + \fB^\mu \ri)  + {2 a_3 \ov \beta} \fw^\mu , \\
\hat j^\mu_o = 3c\hbar \hmu (\hmu \fw^\mu - 2 \fB^\mu)  + 2a_1 \fB^\mu + a_2 \fw^\mu ,
\label{m1}
\end{gather}
and
\ie
\label{po3}
\ell^\mu_o & = 3c\hbar \hmu  \le[\le(1 - {\al \ov 3} \ri) \hmu \fw^\mu  - \le(2 - \al \ri) \fB^\mu \ri]
\cr
& \qquad +  2 a_1 (1 - \al ) \fB^\mu + a_2 \le[(1 - \al ) \fw^\mu -{\al \ov \hmu} \fB^\mu  \ri]  - {2 a_3 \al \ov \hmu}  \fw^\mu  \cr
& = \xi_{\fw} \fw^\mu + \xi_{\fB} \fB^\mu
\fe
with
\be
\xi_\fw = 3c\hbar \hmu^2 \le(1 - {\al \ov 3} \ri) + a_2 (1 - \al ) - {2 a_3 \al \ov \hmu} , \quad
\xi_\fB = - 3c\hbar \hmu (2 - \al ) +  2 a_1 (1 - \al ) -  {a_2 \al \ov \hmu} \ .
\ee

Similarly  $Q$ of~\eqref{jio1} can be written more transparently in the basis of~\eqref{bewb} as
\be \label{newq}
Q = -  c\hbar \le(2 \hmu b \wedge db - \hmu^2 b \wedge dv \ri)
+ {a_1 } b \wedge db + {a_2} v \wedge db + {a_3 } v \wedge dv 
\ee
where we have dropped an exact three-form as
mentioned earlier $Q$ is defined only up to a closed three-form. 





\section{Equilibrium partition function and global gravitational anomalies} \label{sec:par}

In this section we first explain how to obtain the equilibrium partition function from the hydrodynamical effective action.
We discuss two different ways of doing it.
We then apply the procedures to $I_{\rm odd}$ found in {the} last section to obtain the
parity-odd part of the equilibrium partition function. 
We will see that in the absence local anomalies, i.e. $c=0$, all the parity-odd transport terms are connected to {\it global} anomalies. When the underlying theory is only invariant under $\sT$, terms proportional to $a_1, a_2, a_3$ in~\eqref{qo0}--\eqref{po2} are respectively associated with global U(1), mixed gravitational, and gravitational anomalies.
With $\sC \sP \sT$ invaraince, only a global mixed gravitational anomaly is present. This connection also implies that $a_{1,2,3}$ should be proportional to $\hbar$.

\subsection{ Equilibrium partition function from effective action}

We will now describe two methods of obtaining the equilibrium partition function from the effective action
when $\rho_0$ in~\eqref{pager1} is given by the thermal density matrix with an inverse temperature $\beta_0$. By definition the generating functional $W$ of~\eqref{pager1} becomes identically zero when we set the external fields for the two legs to be the same. Nevertheless, as already indicated in~\cite{CGL,CGL1,GL}, the equilibrium partition function can be extracted from the effective action with the help of the dynamical KMS condition. We will again work {to leading order} in small $\hbar$ expansion.

For notational simplicity we will now denote the sources collectively by $\phi_i$ and their corresponding operators $\sO_i$ with index $i$ labelling different operators/components.
In~\cite{CGL} it was shown that a generating functional $W$ satisfying the combined $\Th$ and KMS transformation~\eqref{wkms}
can be ``factorized''  in the stationary limit. That is, when the sources $\phi_{1i}, \phi_{2i}$ are time independent, to leading order in the $a$-field expansion we can write $W$ as
\be \label{1fac}
W [\phi_{1i}, \phi_{2i}] 
=
i \tilde W [\phi_{1i}] - i \tilde W [\phi_{2i}] + \cdots
\ee
where $\cdots$ denotes terms of order $O(a^2)$,  $\tilde W [\phi_i (\vx)]$ is a functional defined on the spatial manifold of the spacetime, and satisfies
\be \label{kio}
\tilde W [\phi_i (\vx) ] = \tilde W [\Theta\phi_i (\vx)]  \ ,
\ee
{where $\Theta$ here should be understood as the extension of (\ref{tg1}) to time-independent field configurations.} Equation~\eqref{1fac} implies that
\be \label{1con2}
 \vev{\sO_{i} (\om =0, \vx)}_\phi  =  {\de \tilde W [\phi_i (\vx)] \ov \de \phi_{i} (\vx)}  \ .
 \ee
Writing the equilibrium partition function $Z$ as
\be
Z = e^{-\beta_0 F}  \ ,
\ee
where $F$ is the free energy, and doing analytic continuation of $\tilde W$ to Euclidean signature,\footnote{See sec. \ref{sec:4b} for an explicit example of the continuation.} from~\eqref{1con2} we can identify $ -\tilde W$ with $\beta_0 F$.

The free energy $F$ (and thus $\tilde W$) should have a local expansion in terms of external sources, as
the equilibrium partition function can be computed by putting the system on a Euclidean manifold with a periodic time circle, which generates a finite gap. As discussed in~\cite{CGL} we can obtain $\tilde W$ from the contact terms in $I_{\rm hydro}$
as follows. One first obtains the source action $I_s$ by setting the dynamical fields in
$I_{\rm hydro}$  to the following equilibrium values

\be \label{equv}
\vp = \vp_a = X_a^\mu =0, \qquad  u^\mu = {1 \ov b} (1, \vec 0), 
\qquad
 \beta = \beta_0 b, \qquad  b = \sqrt{- g_{00}}
\ee
which give
\be
G_{a \mu \nu} = g_{a \mu \nu}, \quad C_{a \mu} = A_{a \mu}  , \quad  B_{\mu} = A_\mu , \quad \hmu = \beta_0 A_0 ,\quad ,
u_0 = - b, \quad u_i = {g_{0i} \ov b}\ .
\ee
All external fields are taken to be time independent.
Then to leading order in the $a$-field expansion
\be
I_s = \tilde W [\phi_1] - \tilde W [\phi_2]  + \cdots
\ee
where $\cdots $ denotes terms of order $O(a^2)$. That $I_s$ is factorizable at this order
is warranted by the dynamical KMS condition.\footnote{In~\cite{CGL} the KMS condition on $I_{\rm hydro}$ was imposed
by requiring $I_s $ to satisfy the combination of $\Th$ and KMS, dubbed the local KMS condition there. In~\cite{CGL1} it was shown the dynamical KMS~\eqref{bakv} and local KMS conditions are equivalent.}

There is also an alternative way to obtain the equilibrium free energy as follows. The dynamical KMS condition~\eqref{bakv} implies that
\be
\tilde \sL= \sL +  \p_\mu V^\mu
\ee
where $\tilde \sL$ 
is defined as  $\tilde \sL = \sL[\Th \tilde \Lam_a, \Th \tilde \Lam_r]$ (see~\eqref{hyi}).
$V^\mu$ can be further expanded in terms of $a$-fields as
\be
V^\mu =i V^\mu_0 + \cdots
\ee
where $V^\mu_0$ contains $r$-fields only. From the discussion of the entropy current in~\cite{GL}, we can then identify\footnote{See equation (3.14) there. The second term $\hat V_1^0$ vanishes in the stationary limit.}
\be \label{part2}
{\int d^{d-1} x \, \sqrt{-g} V_0^0 \bigr|_{\rm eq} = \log Z = - \beta_0 F}
\ee
where $V_0^0 \bigr|_{\rm eq} $ denotes the expression obtained by setting dynamical fields in $V^0_0$ to equilibrium values~\eqref{equv}.

The equivalence of the two methods can be considered as a consequence of equivalence of local KMS condition of~\cite{CGL} and the dynamical KMS condition~\eqref{bakv} as shown in~\cite{CGL1}.
One can readily check that applied to the parity even part of the effective action $I_{\rm even}$ the two methods indeed give the same answers and are equivalent to the results discussed in~\cite{Banerjee:2012iz,Jensen:2012jh}. 

\subsection{Parity-odd equilibrium partition function and global anomalies}\label{sec:4b}

We  now obtain the parity-odd partition function from $I_{\rm odd}$ following the procedures discussed in the previous subsection.
It can be readily checked that the two approaches give the same answers.
The second approach is significantly simpler technically, which we will describe here.
Recall that from our analysis for $\Theta=\mathcal P\mathcal T$ there is no parity-odd contribution to the partition function at first derivative order. The results below are for $\Theta=\mathcal{T}$; to obtain $\Theta=\mathcal{CPT}$ one needs to take $a_{1,3}=0$ together with (\ref{pou}).

From~\eqref{jio},~\eqref{jio1}, and~\eqref{part2} we immediately obtain that
\be \label{w0}
\log Z = \int  \le[  \le(-2 c \hbar \hmu + a_1 \ri) A \wedge dA
+ H_2 u  \wedge dA  +   H_1  u \wedge du \ri] \
\ee
where the integration is over the {\it spatial} manifold with $A \equiv A_i d x^i, u \equiv u_i dx^i$. Using the basis of~\eqref{newq}, equation~\eqref{w0} can be written more transparently as
\be \label{ww0}
\log Z = \int \le[-  c \hbar \beta_0 \le(2 A_0 b \wg d b - A_0^2 b \wg dv \ri) + a_1 b \wg d b + {a_2 \ov \beta_0} v \wg d b
+ { a_3 \ov \beta_0^2} v \wg dv  \ri] \
\ee
where
\be \label{peh}
v = v_i dx^i, \qquad b = b_i dx^i, \qquad v_i  =- \frac{g_{0i}}{ g_{00}},  \qquad b_i =A_i+v_iA_0 \ .
\ee
Equation~\eqref{ww0} precisely agrees with that given in~\cite{Banerjee:2012iz}.

Let us now explore a bit further physical implications of~\eqref{ww0}. The background fields in~\eqref{ww0} are
those for a stationary Lorentzian manifold with
\be \label{bni}
ds^2 = g_{00} \le(dt - v_i dx^i \ri)^2 + a_{ij} dx^i dx^j , \qquad A_\mu dx^\mu  = A_0 dt + A_i dx^i  \
\ee
and $g_{00} < 0$. Note that~\eqref{bni} is preserved by time reparameterizations $t \to t + f (\vec x)$, under which
\be \label{tu1}
v_i \to v_i - \p_i f , \qquad A_i \to A_i + A_0 \p_i f , \qquad b_i \to b_i ,
\ee
and  time-independent $U(1)$ transformations $A_i \to A_i + \p_i \lam (\vec x)$ under which
\be \label{tu2}
v_i \to v_i, \qquad b_i \to b_i + \p_i \lam   \ .
\ee
Below we will refer to~\eqref{tu1} as time $U(1)$ and~\eqref{tu2} as flavor $U(1)$.

The thermal partition function is usually calculated by analytically continuing to Euclidean signature with $t \to - i \tau$ (with $\tau$ on a circle with period $\beta_0$), and the background fields are taken so that they are real in Euclidean signature. We take the Euclidean metric and gauge field to be of form
\begin{align}
ds^2&=g_{00}(d\tau-v_i dx^i)^2+a_{ij}dx^i dx^j\\
A_\mu dx^\mu &= A_0 d\tau+A_i dx^i
\end{align}
Here, $g_{00}> 0$. Thus, under the analytic continutaion $t\rightarrow -i\tau$, we get the replacements
\be
v_i \to  -i v_i, \qquad A_0 \to i A_0
\ee
after which~\eqref{ww0} becomes
\be \label{ww1}
\log Z = \int \le[ -i c \hbar \beta_0 \le(2 A_0 b \wg d b - A_0^2 b \wg dv \ri) + a_1 b \wg d b -i  {a_2 \ov \beta_0} v \wg d b
- { a_3 \ov \beta_0^2} v \wg dv  \ri] \ .
\ee
Note that $\sC \sP \sT$ invariant terms become pure imaginary while the terms proportional to $a_1$ and $a_3$ {\it remain} real.

Now let us consider a system with {\it no} local anomalies, i.e. $c=0$. Then in~\eqref{ww1} we have three Chern-Simons terms, respectively, for flavor $U(1)$, mixed time and flavor $U(1)$, and time $U(1)$. A defining feature
of  Chern-Simons terms is that they are not invariant under ``large'' gauge transformations i.e. those are not connected to the identity. Consider for example the flavor $U(1)$ Chern-Simons term
\be
a_1\int  b \wg d b \ .
\ee
Let us take the spatial manifold to have the topology of $S^1 \times S^2$, where $S^1$ has size $L$. We can choose $b$ to have a monopole configuration on $S^2$, i.e.
\be
\int_{S^2} db = {2 \pi n \ov q} , \qquad n \in \ZZ
\ee
where $q$ is the minimal charge under $U(1)$.

A large gauge transformation of $b_x$ ($x$ is the circle direction)
is
\be
b_x \to b_x + {2 \pi m \ov q  L }, \qquad m \in \ZZ
\ee
we then have~\cite{Poly:1987,Alvarez:1985} 
\be \label{g1}
Z \to e^{8 \pi^2 mn  a_1 \ov q^2} Z  \ .
\ee

Under Kaluza-Klein reduction, $v$ couples to matter as a $U(1)$ gauge
field with minimal ``charge'' ${2 \pi \ov \beta_0}$,thus a large gauge transformation of $v_x$ is
\be \label{poo}
v_x \to v_x + \frac{k \beta_0}{L}, \qquad k \in \ZZ
\ee
we have
\be
Z \to e^{-i   n k {2 \pi \ov q} a_2} Z  \ .
\ee

For the last term in~\eqref{ww1} we need to consider a monopole configuration for $v_i$ on $S^2$,
\be\label{goo}
\int_{S^2} dv = l\beta_0, \qquad l \in \ZZ
\ee
and then under just a large gauge transformation~\eqref{poo} we have
\be \label{g3}
Z \to e^{- 2a_3 kl} Z
\ee
Note in~\eqref{g1} and~\eqref{g3} the partition function transforms by a real number rather than a phase. As mentioned earlier non-vanishing $a_1$ or $a_3$ breaks $\sC \sP \sT$.

Thus we find in the absence of local anomaly, all the anomalous transports are associated with global gauge or gravitational anomalies for putting the system on a Euclidean four-manifold with a thermal time circle.

In the presence of a local anomaly, i.e. $c \neq 0$, then the transport coefficients in~\eqref{m0}--\eqref{po3} are then mixed
among local and global anomalies. The same thing happens to the partition function. But note that $a_1, a_3$ are terms, being real, are not mixed with local anomalies.




Possible connections of the term proportional to $a_2$ with mixed global gravitational anomaly was first hinted in~\cite{Golkar:2012kb} and shown explicitly in~\cite{Golkar:2015oxw,Chowdhury:2016cmh} in some free theory models.



\section{Entropy current} \label{sec:ent}

In this section we obtain the entropy current for a $(3+1)$-dimensional parity-violating fluid by applying~\eqref{nmj}. 
One thing to notice is that the anomalous action~\eqref{ianm} does not have the same structure of the rest
of the action. {At $O(a)$ the latter has the form} (now also including the parity-even part, see~\eqref{jke})
\be \label{infp}
{1 \ov \hbar} \sL_{\rm inv} =  \ha \hat T^{\mu \nu} G_{a \mu \nu}  + \hat J^\mu_0 C_{a \mu} + \cdots \ ,
\ee
{which is the form assumed in \cite{GL}. The fact that $I_{\text{anom}}$ has a different structure does not cause a problem, as $I_{\text{anom}}$ is KMS invariant by itself. We can then simply apply the procedure of~\eqref{nmj} to $I_{\rm inv}$ which will generate an entropy current with non-negative divergence.}

%

Now applying~\eqref{nmj} we find that
\be
S^\mu = V_0^\mu - \hat T^{\mu\nu}\beta_\nu-\hat\mu \mathcal {\hat J}^\mu+ c
\hbar \hat\mu  \ep^{\mu\nu\alpha\beta}B_\nu F_{\alpha\beta} \
\ee
and
\be \label{enc1}
\p_\mu S^\mu = R_{\rm even}  \geq 0
\ee
with $R_{\rm even}$ to be divergence of the entropy current of the parity-even part.\footnote{$R_{\rm even}$ is given explicitly in
equation (5.40) of Sec. V C of~\cite{CGL1}. It was denoted as $Q_2$ there.}
Equation~\eqref{enc1} means that parity-odd part does not contribute to entropy dissipation.

From~\eqref{tot2}--\eqref{jio1}, for the parity-odd part, $V_0^\mu$ is simply the dual of $Q$, giving the following odd-parity contribution to the entropy
\ie
S^\mu_o & =  - 2 u^{(\mu} q^{\nu)}_o \beta_\nu-\hat\mu \hat j_o^\mu  +\left( { a_1 \ov 2} \epsilon^{\mu\nu\alpha\beta}  A_\nu F_{\alpha\beta}+ H_2 \sB^\mu + H_1 \om^\mu \right) \cr
& = {a_1 \hmu^2 +2 a_2 \hmu + 3 a_3  \ov \beta^2} \om^\mu + {2 a_1 \hmu + 2 a_2 \ov \beta} \sB^\mu +
{ a_1 \ov 2} \epsilon^{\mu\nu\alpha\beta}  A_\nu F_{\alpha\beta} \cr
& = {a_1}  \ep^{\mu \nu \lam \rho}  b_\nu \p_\lam b_\rho + 3 a_3  \fw^\mu + 2  a_2 \fB^\mu 
\label{enc}
\fe
where we have dropped a term which is dual to an exact $3$-form. Note that this expression is independent of $c$.
The entropy current in the Landau frame is then given by
\ie
S^\mu_o = -\hat\mu \ell_o^\mu  +\left( { a_1 \ov 2} \epsilon^{\mu\nu\alpha\beta}  A_\nu F_{\alpha\beta}+ H_2 \sB^\mu + H_1 \om^\mu \right)
\fe
which gives
\ie
S_o^\mu &= \left(\frac{2c\hbar\hat{\mu}^3}{\beta^2}(1-\alpha)+\frac{a_1\hat{\mu}^2}{\beta^2}(2\alpha-1)+\frac{2\alpha a_2}{\beta^2}\hat{\mu}+\frac{a_3(1+2\alpha)}{\beta^2}\right)\omega^\mu \cr
&+\left(\frac{3c\hbar\hat{\mu}^2}{\beta}(1-\alpha)+\frac{2a_1\hat{\mu}}{\beta}\alpha+\frac{a_2}{\beta}(1+\alpha)\right)\sB^\mu+\frac{1}{2}a_1 \epsilon^{\mu\nu\alpha\beta}A_\nu F_{\alpha\beta} \ .
\fe
The parts of the expression which involve the anomaly coefficient agree with the Landau frame entropy current given in \cite{Son:2009tf} when $a_1=a_2=a_3=0$. Furthermore, there is also agreement with \cite{Neiman:2010} when $a_1=0$.
After dropping duals of exact three forms, the vector above can be written in the new basis introduced here as
\ie
S_o^\mu &={a_1}  \ep^{\mu \nu \lam \rho}  b_\nu \p_\lam b_\rho+c\hbar\hat{\mu}^2(1-\alpha)(3\fB^\mu-\hat{\mu}\fw^\mu)+a_2\left((1+\alpha)\fB^\mu-(1-\alpha)\hat{\mu}\fw^\mu\right)\\
& +a_3(1+2\alpha)\fw^\mu+2a_1\hat{\mu}(\alpha-1)\fB^\mu \ .
\fe

\section{Parity-violating action in $2+1$-dimension} \label{sec:21}

Let us now consider the action for parity-violating terms in $2+1$-dimension. The procedures
are exactly parallel to those of the $3+1$-dimensional
story. So we will be brief, only giving the main results. We will again work to the level of $\sL_2$ as
defined in~\eqref{altex}. The results below are fully consistent with the constitutive relations
presented in~\cite{Jensen:2011xb} from entropy current analysis and those presented in~\cite{Banerjee:2012iz,Jensen:2012jh} using stationary partition
function.

At $O(a)$ {the hydro Lagrangian has} terms
\be \label{oddac1}
  \ha \hat T^{\mu \nu}_o G_{a \mu \nu}  + \hat J^\mu_o C_{a \mu}  \
\ee
and as usual we can decompose $\hat T^{\mu \nu}_o$ and $\hat J^\mu_o$ as
\bega \label{io1}
\hat T^{\mu \nu}_o = \vep_o u^\mu u^\nu + p_o \De^{\mu \nu} + 2 u^{(\mu} q^{\nu)}_o + \Sig_o^{\mu \nu} , \\
\hat J^\mu_o = n_o u^\mu + j_o^\mu  \
\label{io2}
\end{gather}
where $q^{\nu}_o, j^\mu_o$ and  $\Sig_o^{\mu \nu}$ are transverse to $u^\mu$.
For this purpose let us list all the parity-odd scalars, vectors, and tensors which are
diagonal shift invariant at first derivative order\footnote{Note the identities $\ha \De^\mu{_\nu} \ep^{\nu \lam \rho} F_{\lam \rho} =\ep^{\mu \nu \lam} u_\nu F_{\lam \rho} u^\rho$ and $ \De^\mu{_\nu} \ep^{\nu \lam \rho} \nab_\lam u_\rho = - \ep^{\mu \nu \lam} u_\nu \p u_\lam$. }
\bln
& {\rm scalars:}  \qquad s_1 =  \ep^{\mu \nu \lam} u_\mu \nab_\nu u_\lam , \quad
s_2 = \ha \ep^{\mu \nu \lam} u_\mu F_{\nu \lam} \\
& {\rm vectors:}   \qquad  t_1^\mu =  \ep^{\mu \nu \lam} u_\nu v_{1 \lam}
 , \quad   t_2^\mu = \ep^{\mu \nu \lam} u_\nu v_{2\lam} , \quad
t_3^\mu = \ep^{\mu \nu \lam} u_\nu \p_\lam \beta , \quad  t_4^\mu = \ep^{\mu \nu \lam} u_\nu \p_\lam \hmu \\
& {\rm tensors:}  \qquad  \sig^{\mu \nu}_o = \sig_\lam{^{(\nu}}  \ep^{\mu) \rho \lam} u_\rho
\end{align}
where we have introduced
\bega \label{9de}
 v_{1\mu} = \p u_\mu - \beta^{-1} \De_\mu{^\nu} \p_\nu \beta, \qquad v_{2 \mu} =  \beta^{-1} \De_\mu{^\nu} \nab_\nu \hmu - u^\nu F_{\mu \nu} , \\
  \p \equiv u^\mu \nab_\mu, \qquad \sig^{\mu \nu} \equiv \De^{\mu \lam} \De^{\nu \rho} \le(\nab_\lam u_\rho + \nab_\rho u_\lam -  g_{\lam \rho}
\nab_\al u^\al \ri) \
\end{gather}
We can then expand various quantities in~\eqref{io1}--\eqref{io2} as
\bega \label{op1}
\vep_o = g_1 s_1 + g_2 s_2, \qquad p_o = h_1 s_1 + h_2 s_2, \qquad n_o = f_1 s_1 + f_2 s_2 \\
q^\mu_o = \sum_{i=1}^4 k_i t^\mu_i , \qquad j^\mu_o =  \sum_{i=1}^4 l_i t^\mu_i, \qquad
\Sig_o^{\mu \nu} =  \eta_o \sig^{\mu \nu}_o
\label{op2}
\end{gather}
where all coefficients $g_{1,2}, h_{1,2}, f_{1,2}, k_{1,2,3,4}, l_{1,2,3,4}, \eta_o$ are functions of $\beta, \hmu$.\\
At  $O(a^2)$ the complete action at zero derivative order is
\begin{align}\label{veca0}
- i \mathcal {L}^{(2)}&=\frac 14s_{11}(u^\mu u^\nu G_{a\mu\nu})^2+\frac 14s_{22}(\Delta^{\mu\nu}G_{a\mu\nu})^2-\frac 12 s_{12}u^\mu u^\nu G_{a\mu\nu}\Delta^{\mu\nu}G_{a\mu\nu}\\&+\frac 14 s_{33}(u^\mu C_{a\mu})^2+s_{23}u^\mu C_{a\mu}\Delta^{\alpha\beta} G_{a\alpha\beta}- s_{13}u^\mu u^\nu G_{a\mu\nu} u^\alpha C_{a\alpha}+t(G_{a<\mu\nu>})^2\\
&+r_{11}\Delta^{\alpha\beta}u^\nu u^\mu G_{a\mu\alpha}G_{a\nu\beta}+r_{22}\Delta^{\mu\nu}C_{a\mu}C_{a\nu}+2 r_{12}\Delta^{\mu\nu}u^\alpha G_{a\mu\alpha}C_{a\nu}\label{veca}\\&+r\ep^{\mu\nu\lambda}u_\mu C_{a\nu}u^\rho G_{a\lambda\rho} \ .
\end{align}
Note that only the last term is parity-odd.
Here, angular brackets deonte the transverse symmetric traceless part of the corresponding tensor, e.g.,
\begin{align}\label{angb}
C_{\langle\mu\nu\rangle}&=\Delta_\mu^\alpha\Delta_\nu^\beta\biggl(C_{(\alpha\beta)} -\frac{1}{d-1}\Delta_{\alpha\beta}C^{\gamma\delta}\Delta_{\gamma\delta}\biggr)\ .
\end{align}
Non-negativity of the imaginary part of the action, eq. (\ref{keyp1}), leads to various constraints among the coefficients of $\mathcal L^{(2)}$. The constraints on the parity even part (\ref{veca0})-(\ref{veca}) were analyzed in detail in~\cite{CGL}. Among other constraints we have
\be
r_{11}, r_{22} > 0, \qquad  r_{11}r_{22}-r_{12}^2 \geq 0 \ .
\ee
When the parity-odd coefficient $r$ is nonzero, the second inequality of the above becomes
\be \label{jnn}
r_{11}r_{22}-r_{12}^2 \geq {r^2 \ov 4} \ .
\ee


To summarize, to level $\sL_2$ the parity-odd action can be written as
\be
 \sL_{\rm odd} = \ha ( \vep_o u^\mu u^\nu + p_o \De^{\mu \nu} + 2 u^{(\mu} q^{\nu)}_o + \Sig_o^{\mu \nu} ) G_{a \mu \nu}
 + (n_o u^\mu + j_o^\mu ) C_{a \mu} + i \, r \,  \ep^{\mu \nu \lam} u_\mu C_{a \nu} G_{a \lam \rho} u^\rho  \ .
 \ee
Using field redefinitions one can write $\sL_{\rm odd}$ as (see Sec. VI of~\cite{CGL1} for details)
\be \label{lanac1}
\sL_{\rm odd} =   \ha \th_o \Delta^{\mu\nu}G_{a\mu\nu}
+ \ell^\mu_o \De_\mu^\nu C_{a\nu}
\ee
with frame independent quantities $\th_o, \ell_o^\mu$ defined by
\be\label{fde}
\th_o = p_o -   \vep_o  \p_\vep p_0 -  n_o \p_n \ep_0 , \qquad \ell_o^\mu =  j^\mu_o -{ n_0  \ov \varepsilon_0+p_0 } q^\mu_o
\ee
where $\ep_0, p_0, n_0$ are respectively zeroth order energy, pressure and charge densities.
Note that the coefficient $r$ can be defined away using field redefinitions, so~\eqref{jnn} does not lead to new constraints
on transport coefficients.

The outcome of the dynamical KMS condition~\eqref{bakv} again depends very much on the choice $\Th$, which we
will discuss separately.

\subsection{$\Th = \sT$}

In this case, we find all coefficients in~\eqref{op1}--\eqref{op2} are zero, except for $k_2$ and $l_1$ which satisfy
the relation
\begin{align}
-k_2=l_1=\frac{1}{2}\beta r  \ .
\end{align}
The full parity-odd action to level $\sL_2$ can then be
written as
\be
\sL_{\rm odd}  = -\frac{\beta r}{4} (u^\mu t_2^\nu+u^\nu t_2^\mu) G_{a \mu \nu}
+ \frac{\beta r}{2}  t_1^\mu C_{a \mu} +  i \, r  \ep^{\mu \nu \lam} u_\mu C_{a \nu} G_{a \lam \rho} u^\rho \ .
\ee
The above Lagrangian satisfies
\be \label{enr}
\tilde \sL = \sL
\ee
which can be seen by noting the relation
\be
\hat{T}_o^{\mu\nu}\nabla_\mu\beta_\nu+\hat{J}_o^\mu(\partial_\mu\hat{\mu}- \beta^\alpha F_{\mu \alpha}) =-r\beta^2\epsilon^{\mu\nu\rho}u_\mu v_{2\nu}v_{1\rho} \ .
\ee

Due to~\eqref{enr}, there is no parity-odd contribution to the thermal partition function to first derivative order.
The entropy current is given by
\be
S^\mu= p\beta^\mu -T^{\mu\nu}\beta_\nu-\hat{\mu}J^\mu
\ee
where $p, T^{\mu \nu}, J^\mu$ also include the parity-even part, and
\be \label{hhe}
\nabla_\mu S^\mu
=R_{\rm even} +r\beta^2\epsilon^{\mu\nu\rho}u_\mu v_{2\nu}v_{1\rho}
\ee
where $R_{\rm even} $ is the parity-even expression. 
Note that the second term in the right hand side of~\eqref{hhe} vanishes by ideal fluid equation of motion
\begin{align}
\label{ideq}
v_{1\mu}&=-\frac{n_0}{\epsilon_0+p_0}v_{2\mu} \ .
\end{align}

\subsection{$\Th = \sT \sP$}

The dynamical KMS condition implies that the coefficients in~\eqref{op1}--\eqref{op2} should satisfy
\bega
h_1= h_2= r = 0,  \quad  k_2=l_1 \\
%
g_1 = \beta k_3 , \quad - f_1 = \beta k_4 , \quad
g_2 = \beta l_3 , \quad  - f_2 = \beta l_4 \label{pen1} \\
\p_\beta (\beta l_4) = \p_\hmu (\beta l_3) , \qquad
l_3+k_4 = \partial_{\hat{\mu}}(\beta k_3)- \partial_{\beta}(\beta k_4)  \ .
\label{pen2}
\end{gather}
The first equation of~\eqref{pen2} implies that there exists a function $Y$ such that
\be \label{pen3}
 \beta l_3=\frac{\partial Y}{\partial\beta} , \qquad \beta l_4 =\frac{\partial Y}{\partial\hat{\mu}}
 \ee
while the second equation of~\eqref{pen2} can be further written as
\be
\partial_{\beta}(\beta^2 k_4)+\beta l_3 =\partial_{\hat{\mu}}(\beta^2 k_3)
\ee
which upon using~\eqref{pen3} implies that there exists a function $X$ such that
\be \label{pen4}
\beta^2 k_3 =\partial_\beta X , \qquad \beta^2 k_4+Y  =\partial_{\hmu} X  \ .
\ee
$k_1,l_2, \eta_o$ are unconstrained. Thus there are altogether six
 independent functions of $\hmu$ and $\beta$: $X, Y, k_1, l_1, l_2, \eta_o$.

Applying the above relations to~\eqref{op1}--\eqref{op2} we then have
\bega \label{tt1}
\hat T^{\mu \nu}_o = \le({1 \ov \beta} \p_\beta X s_1 + \p_\beta Y s_2\ri) u^\mu u^\nu +
2   u^{(\mu}  q^{\nu)}_o 
 + \eta_o \sig^{\mu \nu}_o  , \\
 \label{tt2}
 q_o^\mu =  k_1 t^\mu_1+ l_1   t^\mu_2+ {1 \ov \beta^2} \p_\beta X  t^\mu_3+ {1 \ov \beta^2} (\p_\hmu X - Y)   t^\mu_4 , \\
\hat J^\mu_o = \le({1 \ov \beta} (Y - \p_\hmu X) s_1 - \p_\hmu Y s_2 \ri) u^\mu + l_1 t^\mu_1+ l_2   t^\mu_2+ {1 \ov \beta} \p_\beta Y  t^\mu_3+ {1 \ov \beta} \p_\hmu Y   t^\mu_4 \ .
\label{tt3}
\end{gather}
It can be checked that the above expressions satisfy
\be
\hat{T}_o^{\mu\nu}\nabla_\mu\beta_\nu+\hat{J}_o^\mu(\partial_\mu\hat{\mu}+\beta^\alpha F_{\alpha\mu}) =
\nab_\mu V^\mu_0\label{relat1}
\ee
with
\be \label{pen42}
V_0^\mu = \frac{1}{\beta}\partial_\beta X t_3^\mu+\frac{1}{\beta}(\partial_{\hat{\mu}} X- Y ) t_4^\mu +{Y \ov 2} \epsilon^{\mu\nu\rho}F_{\nu\rho}
\ee
which gives
\be \label{pen41}
\tilde{\mathcal{L}}_{\rm odd}-\mathcal{L}_{\rm odd} 
= i \nab_\mu V^\mu_0 \ .
\ee

The entropy current can then be obtained as
\begin{align}
S^\mu&=p\beta^\mu+\frac{1}{\beta}(\partial_\beta X) t_3^\mu+\frac{1}{\beta}(\partial_{\hat{\mu}}X-Y) t_4^\mu+\frac{Y}{2}\epsilon^{\mu\nu\rho}F_{\nu\rho}-T^{\mu\nu}\beta_\nu-\hat{\mu}J^\mu
\end{align}
with
\be \label{pen44}
\nab_\mu S^\mu = R_{\rm even} \ .
\ee


To compare with \cite{Jensen:2011xb}, note that we need to first add the total derivative with zero divergence $-\nabla_\nu(\epsilon^{\mu\nu\rho}u_\rho\tilde{\nu}_5)$ to their expression of the entropy current. This has the consequence of redefining
\begin{align}
\tilde{\nu}_1\rightarrow\tilde{\nu}_1+\partial_T\tilde{\nu}_5\\
\tilde{\nu}_3\rightarrow\tilde{\nu}_3+\partial_{\hat{\mu}}\tilde{\nu}_5
\end{align}
in their expressions. Further comparing corresponding terms, we find Eq.(3.22),Eq.(3.23) and Eq.(3.24) in \cite{Jensen:2011xb} are reproduced if we make the identifications
\begin{align}
X&=\frac{\mathcal{M}_\Omega}{T^2}+\int^T\frac{f_\Omega(x)}{x^3}dx\label{comp1}\\
Y&=\tilde{\nu}_4=\beta\mathcal{M}_B\label{comp2}
\end{align}

For stationary sources~\eqref{bni}, the thermal partition function is obtained from the zeroth component of $V_0^\mu$ with dynamical fields set to their equilibrium values. We find that
\be \label{pen45}
\log Z =\int d^2 x \, \sqrt{-g} \le( {1 \ov \beta_0} \epsilon^{ij}\partial_i v_j\biggl(X-\beta_0 A_0 Y \biggr)+Y \epsilon^{ij}\partial_i b_j \ri)
\ee
where $b_i$ is as defined in~\eqref{peh}. The above expression of the partition function agrees with \cite{Banerjee:2012iz,Jensen:2012jh}.


Finally let us note that the frame independent quantities~\eqref{fde} can be written as
\bega \label{tho}
\th_o ={\chi}_\Omega s_1+ {\chi}_B s_2, \\
 \chi_\Om = -\partial_{\epsilon_0}p_0\biggl(\frac{1}{\beta}\partial_\beta X \biggr)+\partial_{n_0}p_0\frac{1}{\beta}(\partial_{\hat{\mu}}X -Y ), \\
 \chi_B = -\partial_{\epsilon_0}p_0 \partial_\beta Y +\partial_{n_0}p_0 \partial_{\hat{\mu}}Y
\end{gather}
and
\bega
\ell^\mu_o = - \tilde{\sigma} t_2^\mu + \tilde{\chi}_E\tilde{E}^\mu- \frac{ \tilde{\chi}_T}{\beta^2}t_3^\mu , 
\qquad \tilde{E}^\mu  \equiv \beta^{-1}t_4^\mu-t_2^\mu = \ep^{\mu \nu \lam} u_\nu F_{\lam \rho} u^\rho \\
\tilde{\chi}_E  =\partial_{\hat{\mu}}Y -\frac{n_0}{\beta(\epsilon_0+p_0)}(\partial_{\hat{\mu}}X -Y) ,\\
\tilde{\chi}_T =-\beta\partial_\beta Y +\frac{n_0}{\epsilon_0+p_0}\partial_\beta X , \label{tho00} \\
-\tilde{\sigma}
=l_2-\frac{2n_0}{\epsilon_0+p_0}l_1+\biggl(\frac{n_0}{\epsilon_0+p_0}\biggr)^2k_1 
+ \tilde \chi_E \ .
\label{tho1}
\end{gather}
Note that in the above expressions we have used ideal fluid equation~\eqref{ideq} which makes $t_1^\mu$ and $t_2^\mu$ equivalent. As a result the number of independent functions reduce to four: $ X, Y, \tilde \sig, \eta_o$.

\subsection{$\Th = \sC \sP \sT$}

Dynamical KMS invariance requires that  $k_1, k_3, l_2, l_4, g_1, f_2, r, \eta_o$ be even functions of $\hmu$, while $k_4, l_3, g_2,  f_1$ be  odd functions of $\hmu$,
\bega
h_1 = h_2 =0, \qquad k_2-l_1 = -\beta r  , \\
- k_2(-\hat{\mu},\beta) =k_2(\hat{\mu},\beta)+\beta r(\hat{\mu},\beta) , \qquad
 l_1(-\hat{\mu},\beta)= - l_1(\hat{\mu},\beta)+\beta r(\hat{\mu},\beta) ,
\end{gather}
and~\eqref{pen1}--\eqref{pen4}, except that now  $X$ should be an even function of $\hat{\mu}$ and $Y$ should be odd.
Thus equations~\eqref{tt1} and~\eqref{tt3} are unchanged while~\eqref{tt2} should be modified to
\be
q_o^\mu =  k_1 t^\mu_1+ (l_1 - \beta r)  t^\mu_2+ {1 \ov \beta^2} \p_\beta X  t^\mu_3+ {1 \ov \beta^2} (\p_\hmu X - Y)   t^\mu_4  \ .
\ee

Equations~\eqref{pen41}--\eqref{pen45} still apply except that for~\eqref{pen44} the covariant derivative of $S^\mu$ now yields
\be
\nabla_\mu S^\mu = R_{\rm even} + r\beta^2\epsilon^{\mu\nu\rho}u_\mu v_{2\rho}v_{1\nu}
\ee
with the second term on right hand side again vanishing from ideal fluid equation of motion~\eqref{ideq}.

Eqs. \eqref{tho}--\eqref{tho00} are unmodified, while

\be -\tilde{\sigma}
=l_2-\frac{2n_0}{\epsilon_0+p_0}l_1+\biggl(\frac{n_0}{\epsilon_0+p_0}\biggr)^2k_1 
+\frac{\beta n_0}{\ep_0+p_0}r+ \tilde \chi_E \ .\ee




\section*{Acknowledgements}
We would like to thank Umut Gursoy, Kristan Jensen, Zohar Komargodski, Karl Landsteiner, Domingo Gallegos Pazos,  Andrey Sadofyev, Savdeep Sethi,
Dam Thanh Son, Juven Wang, Xiao-Gang Wen, and Amos Yarom for discussions.
This work is supported by the Office of High Energy Physics of U.S. Department of Energy under grant Contract Number  DE-SC0012567. P. G. was supported by a Leo Kadanoff Fellowship.

\appendix

\section{Explicit expressions for various discrete transformations} \label{app:cpt}

In this Appendix we list transformations of various tensors under
various discrete symmetries. They are important for obtaining the explicit forms of dynamical KMS transformations~\eqref{hyi} and~\eqref{si2}--\eqref{si22} of various tensors.
For notational simplicity we have suppressed the transformations of the arguments of all the functions, which are given in the first line of each table.

\medskip

\begin{center}
\begin{tabular}{ |p{2cm}||p{3cm}|p{2cm}|p{2cm}|  }
 \hline
 \multicolumn{4}{|c|}{Discrete transformations in 3+1-dimension} \\
  \hline
 & $\sT$ &  $\sP \sT$ & $\sC \sP \sT$ \\
  \hline
 $x^\mu$   & $(-x^0,  x^i)$    & $- (x^0,  x^i)$    &  $- (x^0,  x^i)$ \\
 $u^\mu$   & $(u^0, - u^i)$    & $(u^0,  u^i)$    &  $(u^0,  u^i)$ \\
 $\om^\mu$   & $(\om^0, - \om^i)$    & $-(\om^0,  \om^i)$    &  $-(\om^0,  \om^i)$ \\
 $A_\mu$ &   $(A_0, - A_i)$     &  $(A_0,  A_i)$   &$- (A_0,  A_i)$\\
 $\sB^\mu$&  $(\sB_0, - \sB_i)$     &  $- (\sB_0,  \sB_i)$   &$(\sB_0,  \sB_i)$\\
$\p_\mu$ &$(-\p_0,  \p_i)$ & $- (\p_0, \p_i)$&  $- (\p_0, \p_i)$ \\
 $\p = u^\mu \p_\mu$    & $- \p$  & $- \p$ &  $- \p$ \\
  $\hmu$ &   $\hmu$   &  $\hmu$  & $- \hmu$  \\
 $g_{\mu \nu}$ & $(g_{00}, - g_{0i}, g_{ij})$ & $ g_{\mu \nu}$
 & $g_{\mu  \nu}$  \\
  $\vp$ &    $- \vp$   &  $- \vp$  & $\vp$  \\
 \hline
\end{tabular}
\end{center}


\begin{tabular}{ |p{2cm}||p{3cm}|p{6cm}|p{6cm}|  }
 \hline
 \multicolumn{4}{|c|}{Discrete transformations in 2+1-dimension} \\
  \hline
 & $\sT$ &  $\sP \sT$ & $\sC \sP \sT$ \\
 \hline
 $x^\mu$   & $(-x^0, x^i)$    & $(- x^0,  -x^1,  x^2)$    &  $(-x^0,  -x^1, x^2)$ \\
 $u^\mu$   & $(u^0, - u^i)$    & $(u^0,  u^1, - u^2)$    &  $(u^0,  u^1, - u^2)$ \\
 $A_\mu$ &   $(A_0, - A_i)$     &  $(A_0,  A_1, - A_2)$   &$(-A_0,  -A_1,  A_2)$\\
$\p_\mu$ &$(-\p_0,  \p_i)$ & $(-\p_0, -\p_1,  \p_2)$&  $(-\p_0, -\p_1,  \p_2)$ \\
 $\p = u^\mu \p_\mu$    & $- \p$  & $- \p$ &  $- \p$ \\
  $\hmu$ &   $\hmu$   &  $\hmu$  & $- \hmu$  \\
 $v_1^\mu$ & $(-v_{1}^0, v_{1}^i)$  & $(-v_{1}^0, -v_{1}^1, v_{1}^2 )$  &$(-v_{1}^0, -v_{1}^1, v_{1}^2 )$ \\
 $v_2^\mu$ & $(-v_{2}^0, v_{2}^i)$  & $(-v_{2}^0, -v_{2}^1, v_{2}^2 )$  & $(v_{2}^0, v_{2}^1, -v_{2}^2 )$\\
 $g_{\mu \nu}$ & $(g_{00}, - g_{0i}, g_{ij})$ & $(g_{00}, g_{01}, -g_{02}, - g_{12}, g_{11}, g_{22})$
 & $(g_{00}, g_{01}, -g_{02}, - g_{12}, g_{11}, g_{22})$  \\
 $s_1$ & $- s_1$ & $s_1$ & $s_1$ \\
  $s_2$ & $- s_2$ & $s_2$ & $- s_2$ \\
  $t_\al^\mu, \al=1,3$ & $(-t_{\al}^0, t_{\al}^i)$  & $(t_{\al}^0, t_{\al}^1, - t_{\al}^2 )$  &$(t_{\al}^0, t_{\al}^1, - t_{\al}^2 )$ \\
  $t_\al^\mu, \al=2,4$ & $(-t_{\al}^0, t_{\al}^i)$  & $( t_{\al}^0, t_{\al}^1, - t_{\al}^2 )$  &$(-t_{\al}^0, -t_{\al}^1,  t_{\al}^2 )$ \\
  $\sig_o^{\mu \nu}$ & $(-\sig_o^{00}, \sig_o^{0i}, -\sig_o^{ij})$ & $(\sig_o^{00},  \sig_o^{01},- \sig_o^{02}, - \sig_o^{12}, \sig_o^{11}, \sig_o^{22})$
 & $( \sig_o^{00}, \sig_o^{01}, -\sig_o^{02}, - \sig_o^{12},  \sig_o^{11},  \sig_o^{22})$   \\
 \hline
\end{tabular}
Note that in all cases  $C_{a \mu}, J^\mu$  transform as $A_\mu$, while $, T^{\mu \nu}, g_{a \mu \nu}$ have the same transformations as $g_{\mu \nu}$, and $\sig^{\mu \nu}$ transforms the same as $g_{\mu \nu}$ but with an overall minus sign.
$\om^\mu$ and $\sB^\mu$ are defined below~\eqref{vortcurrent}.

\section{Some useful formulae} \label{app:a}

In this Appendix we give some useful formulae used in deriving equations such as~\eqref{jio} and~\eqref{tot}.

We first note an identity in $(3+1)$-dimension
\be \label{uid}
V_\mu\epsilon^{\alpha\beta\gamma\delta}F_{\alpha\beta} G_{\gamma\delta}  =
- 2 \epsilon^{\alpha\beta\gamma\delta} F_{\mu \al} V_\beta G_{\gamma\delta} -
 2 \epsilon^{\alpha\beta\gamma\delta} G_{\mu \al} V_\beta F_{\gamma\delta}  \
\ee
which can be written in differential forms as
\be
\xi \cdot V F \wedge G = - (\xi \cdot F) \wedge V \wedge G -  (\xi \cdot G) \wedge V \wedge F \ .
\ee
 where $\xi$ is a vector field, $F,G$ are two-forms, and $V$ is a one-form.
As an example, given $u \equiv u_\mu dx^\mu$, $w = d u$,  and $\beta^\mu = \beta u^\mu$, we then have
\be
 \beta F \wedge w = (\beta \cdot F)
\wedge u \wedge w  +  (\beta \cdot w) \wedge u \wedge F \ .
\ee
It is also useful to recall that for a differential form $\lam$ and a vector field $\xi$
\be
d (\xi \cdot \lam) = \sL_\xi \lam - \xi \cdot d \lam \ .
\ee
It then follows that  for some vector $v^\mu$
\be
\int \sL_{v} \vp F \wedge F = -2 \int  \vp \, F \wedge \sL_{v} F  + 2 \int d (\vp F \wedge (v \cdot F))
\ee
which can be used to derive~\eqref{jio}.

To see~\eqref{tot}, we note that:
\bea
- c \hbar  \int d^4x \, \sqrt{-g}\ep^{\mu \nu \lam \rho}   F_{ \nu \lam} B_\rho  \sL_{ \beta} B_\mu
&= & - 2 c \hbar \int \sL_\beta B \wedge B  \wedge  F \cr
&=&  - 2 c \hbar \int \le(d \hmu  + \beta \cdot F \ri)  \wedge B  \wedge  F \cr
&=& 3 c \hbar \int \hmu F \wedge F -  2 c \hbar \int  d (\hmu B \wedge F)
\eea
\bea
\int d^4x \sqrt{-g} \, \hat j_o^\mu \sL_{ \beta} B_{\mu}  &= & \int \sL_\beta B \wedge u \wedge (h_1 du + h_2 F)  \cr
&=&  \int  \le(d \hmu  + \beta \cdot F \ri)  \wedge u \wedge (h_1 du + h_2 F) \cr
&=& \int \le[ d \hmu   \wedge u \wedge (h_1 du + h_2 F) +h_1  (\beta \cdot F) \wedge u \wedge  du  + {\beta h_2 \ov 2} F \wedge F \ri]
\eea
\bea
&& \ha \int d^4 x \, \sqrt{-g} \,  T^{\mu \nu}_o \sL_{\beta} g_{\mu \nu} = 
\int d^4 x \, \sqrt{-g} \, \le( (\beta \cdot du)_\mu  - \nab_\mu  \beta \ri) q^\mu_o \cr
 && =
\int \le[{ g_1\beta \ov 2}  du \wedge du +  g_2 (\beta \cdot du) \wedge u \wedge F
 - d \beta \wedge u \wedge (g_1 du  + g_2 F) \ri] \ .
\eea

For $(2+1)$-dimension we have
\be
V_\mu \ep^{\al \beta \ga} W_\al G_{\beta \ga} = W_\mu \ep^{\al \beta \ga} V_\al G_{\beta \ga}
-2 \ep^{\al \beta \ga} G_{\mu \al} V_\beta W_\ga
\ee
or in differential form
\be
(\xi \cdot V) W \wedge G = (\xi \cdot W) V \wedge G -  (\xi \cdot G) \wedge V \wedge W
\ee
where $W$ is a one-form and $G$ a two-form.
Here are two examples:
\bega
\ep^{\mu \nu \lam} u_\nu F_{\lam \rho} u^\rho = \ha \De^\mu{_\nu} \ep^{\nu \lam \rho} F_{\lam \rho} \\
- \ep^{\mu \nu \lam} u_\nu \p u_\lam = \ep^{\mu \nu \lam} u_\nu w_{\lam \rho} u^\rho = \ha \De^\mu{_\nu} \ep^{\nu \lam \rho} w_{\lam \rho}
\end{gather}
where $w = du$.

\end{document}